\documentclass[fleqn,usenatbib]{mnras}

\usepackage{newtxtext,newtxmath}

\usepackage[T1]{fontenc}
\usepackage{ae,aecompl}
\usepackage{float}

\usepackage{amsmath,amstext}
\usepackage[figure,figure*]{hypcap}

\usepackage{graphicx}
\usepackage{amssymb}

\title[GASP XV. Properties of the Jellyfish Galaxy JO201]{GASP. XV. A MUSE View of Extreme Ram-Pressure Stripping along the Line of Sight:\\ Physical properties of the Jellyfish Galaxy JO201}

\author[C. Bellhouse et al.]{
Callum Bellhouse$^{1}$\thanks{E-mail: callumb@star.sr.bham.ac.uk},
Y.~L. Jaff\'e$^{2}$,
S.~L. McGee$^{1}$,
B.~M. Poggianti$^{3}$,\newauthor
R. Smith$^{4}$,
S. Tonnesen$^{5}$,
J. Fritz$^{6}$,
G. K. T. Hau$^{7}$,
M. Gullieuszik$^{3}$,\newauthor
B. Vulcani$^{3}$,
G. Fasano$^{3}$,
A. Moretti$^{3}$,
K. George$^{8,9}$,
D. Bettoni$^{3}$,\newauthor
M. D'Onofrio$^{10,3}$,
A. Omizzolo$^{3,11}$,
Y.-K. Sheen$^{4}$
\\
$^{1}$University of Birmingham School of Physics and Astronomy, Edgbaston, Birmingham, United Kingdom\\
$^{2}$Instituto de F\'isica y Astronom\'ia, Universidad de Valpara\'iso, Avda. Gran Breta\~na 1111 Valpara\'iso, Chile\\
$^{3}$INAF - Astronomical Observatory of Padova, vicolo dell'Osservatorio 5, I-35122 Padova, Italy\\
$^{4}$Korea Astronomy and Space Science Institute, 776, Daedeokdae-ro, Yuseong-gu, Daejeon, 34055, Korea\\
$^{5}$Center for Computational Astrophysics, Flatiron Institute, 162 5th Ave, New York, NY 10010, USA\\
$^{6}$Instituto de Radioastronom\'ia y Astrof\'isica, UNAM, Campus Morelia, A.P. 3-72, C.P. 58089, Mexico \\
$^{7}$European Southern Observatory, Alonso de Cordova 3107, Vitacura, Casilla 19001, Santiago de Chile, Chile\\
$^{8}$Indian Institute of Astrophysics, Koramangala II Block, Bangalore, India\\
$^{9}$Department of Physics, Christ University, Bangalore, India\\
$^{10}$Dipartimento di Fisica \& Astronomia ``Galileo Galilei'', Universit\`a di Padova, vicolo dell' Osservatorio 3, IT 35122, Padova, Italy\\
$^{11}$Specola Vaticana, 00120, Vatican City State\\
}

\date{Accepted XXX. Received YYY; in original form ZZZ}

\pubyear{2018}

\begin{document}
\label{firstpage}
\pagerange{\pageref{firstpage}--\pageref{lastpage}}
\maketitle

\begin{abstract}

We present a study of the physical properties of JO201, a unique disk galaxy with extended tails undergoing extreme ram-pressure stripping as it moves through the massive cluster Abell 85 at supersonic speeds mostly along the line of sight. JO201 was observed with MUSE as part of the GASP programme. In a previous paper (GASP II) we studied the stellar and gas kinematics. In this paper we present emission-line ratios, gas-phase metallicities and ages of the stellar populations across the galaxy disk and tails. We find that while the emission at the core of the galaxy is dominated by an active galactic nucleus (AGN), the disk is composed of star-forming knots surrounded by excited diffuse gas. The collection of star-forming knots presents a metallicity gradient steadily decreasing from the centre of the galaxy outwards, and the ages of the stars across the galaxy show that the tails formed $\lesssim 10^9$ yr ago. This result is consistent with an estimate of the stripping timescale ($\sim 1$~Gyr), obtained from a toy orbital model. 
Overall, our results independently and consistently support a scenario in which a recent or ongoing event of intense ram-pressure stripping acting from the outer disk inwards, causes removal and compression of gas, thus altering the AGN and star-formation activity within and around the galaxy. 

\end{abstract}



\section{Introduction}

Understanding the processes which alter the gas content of galaxies is crucial to the study of galaxy evolution, since gas is a fundamental driver of the formation of stars. Many of these processes depend strongly on the environment in which the galaxy resides, as evidenced by the correlation of galaxy morphology and colour with environment and galaxy mass \citep[e.g.][]{1980ApJ...236..351D,2006MNRAS.373..469B,2010ApJ...721..193P,Fasano2015,Vulcani2015}.
Denser environments such as groups and clusters can promote gravitational and hydrodynamical encounters which can drastically alter the morphology and composition of a galaxy.

In particular, ram pressure stripping \citep[RPS; ][]{1972ApJ...176....1G} is a
process which can remove the gas from the disk of a galaxy as it plunges through the dense central region of a cluster, experiencing a drag force from the hot intra-cluster medium (ICM) acting upon its cool gas component. If the drag force is sufficiently large to overcome the restoring gravitational force of the galaxy, the gas component is stripped. As the galaxy moves at higher speeds or through denser regions, the stripping intensifies and the amount of gas loss increases.
Various studies have found RPS to strongly influence the evolution of galaxies in clusters \citep{2004ApJ...613..866K}. The most compelling evidence has come from observations of neutral atomic Hydrogen (HI), which reveal long tails of stripped gas, and truncated gas profiles in galaxies falling into massive clusters \citep[][]{Haynes1984,Cayatte1990,2004AJ....127.3361K,2009AJ....138.1741C,Cortese2010,2001ApJ...561..708V,2015MNRAS.448.1715J,Yoon2017} and even lower-mass groups \citep{Rasmussen2006,Rasmussen2008,Jaffe2012,Hess2013}.

Narrow-band imaging centered on the ionised H$\alpha$ emission has also proven to be an excellent tool to study gas stripping phenomena and the subsequent evolution of galaxies \citep{Gavazzi2002a}. In particular, the H$\alpha$ emission allows the study of star formation activity in the ram-pressure interaction, which can result from the compression of the gas present in the disk, as well as the collapse of stripped gas in the tails \citep{Kenney1999,Yoshida2008,Kenney2014,Boselli2018,Fossatti2018,Boselli2018b}.

Galaxies with clear signs of gas stripping in the optical or UV have been named ``jellyfish galaxies", owing to their ``tails" of diffuse gas where there are often knots of recent star formation. 
%
%
Since jellyfish galaxies represent a transition period of galaxies prior to quenching in the cluster environment, they have been the subjects of many studies
\citep{Boselli2016,Fumagalli2014,2016MNRAS.455.2028F}. However, it was not until recently that the first large catalogues of optically-selected jellyfish galaxies at low \citep{Poggianti2016} and intermediate \citep{McPartland2016} redshifts have been produced.


Further advancing the study of gas stripping processes has been the advent of integral field spectrographs (IFS) such as the multi-unit spectroscopic explorer (MUSE) at the very large telescope (VLT). IFS observations allow the spatial distribution of properties to be measured, giving a far greater insight into the kinematic and emission properties \citep[e.g.][]{Fumagalli2014,2016MNRAS.455.2028F}.


GAs Stripping Phenomena in galaxies with MUSE \citep[GASP][]{Poggianti2017}, is a spatially-resolved spectroscopic survey of the gas-removal processes in a sample of 114 disk galaxies 
at redshifts $0.04 < z < 0.07$.
The galaxy coverage in GASP resulting from the $1'\times1'$ field of view of MUSE amounts to $50-100\mathrm{kpc}$ from the galaxy centre, crucial for analysing the diffuse gas tails and extraplanar features of each galaxy. 
The GASP targets were drawn from the jellyfish candidate sample of \citet{Poggianti2016} and were chosen to  span a wide range of galaxy masses ($10^{9.2}-10^{11.5}\mathrm{M}_\odot$) and environments (dark matter haloes range from $10^{11}-10^{15.5}\mathrm{M}_\odot$). 

%

From all the GASP sample, the jellyfish galaxy JO201 is one of the most spectacular and complex examples of ram-pressure stripping at play, occupying a seldom explored region of position vs. velocity phase space of cluster galaxies, with an extreme line of sight velocity of 3363.7km~s$^{-1}$ and a very small projected distance to the centre (360~kpc) of the massive ($\mathrm{M}_{200}=1.58\times10^{15}$) cluster A85. This paper adds JO201 to the small sample of jellyfish galaxies studied in depth to date, with the aim of stretching the RPS parameter space explored so far to its limits.  


In \citet[][GASP II]{Bellhouse2017}, we presented a study of the environment of JO201 and the kinematic properties of its $\mathrm{H}\alpha$ emitting gas. 
We found that JO201 is moving at supersonic speeds through the core of the massive cluster Abell 85 (A85) mostly along the line of sight, unlike the other jellyfish galaxies studied to date \citep{Gullieuszik2017,Poggianti2017,Fumagalli2014}. The GASP MUSE data as well as follow-up multi-wavelength observations revealed a largely undisturbed stellar disk accompanied by a disturbed, asymmetric and extended gas component with in-situ star-formation, suggestive of ongoing intense outside-in ram-pressure stripping along the line-of-sight (see Section~\ref{subsec:jo201} for details). 

In this paper, we present an in-depth analysis of the physical properties of JO201, including gas ionisation from emission-line ratio diagnostics, 
metallicity,  star-formation rates, and ages of the stellar populations across the galaxy and its stripped tails. At the end, we compare all the derived quantities for this peculiar galaxy with a simple model of its evolution within the cluster to construct a comprehensive view of its stripping history. 



The paper is structured as follows. Section \ref{sec:data} summarises the properties of JO201 and the available MUSE data from the GASP survey. 
In section \ref{sec:elfits}, we describe the steps to analyse the datacubes and fit emission lines from the spectra.
In section \ref{sec:analysis} we calculate the gas properties from the emission line parameters.
%
Section \ref{sec:knots} focuses on the knots visible in the disk and tails.
%

Section \ref{sec:sf} covers the star-formation rates across the galaxy, measured using the H$\alpha$ emission.

In section \ref{sec:stellar}, we explore the stellar population, 
comparing the star-formation rate surface density across the history of JO201's interaction with the cluster environment.
%
%
In Section \ref{timescale} we use parameters of the galaxy and its host cluster to explore the stripping history of JO201.
%
Finally in section \ref{sec:conclusions} we summarise the findings and draw conclusions, piecing together the journey of JO201 as it bullets through A85.

Throughout this paper we adopt a Chabrier initial mass function \citep[IMF;][]{Chabrier2003}, and a concordance $\Lambda$CDM cosmology of $\Omega_\mathrm{M}=0.3$, $\Omega_\Lambda=0.7$, $\mathrm{H}_0=70\mathrm{km}\,\mathrm{s}^{-1}\mathrm{Mpc}^{-1}$. 

\section{Data}\label{sec:data}

This study exploits the MUSE observations presented in  GASP II (Sec. 3.1).

In this section we summarise the main findings of previous studies of JO201 (Sec.~\ref{subsec:jo201}), and the GASP MUSE 
observations, data reduction and analysis utilised  (Sec.~\ref{sec:MUSE}), highlighting any differences between the present study and GASP II.  

\begin{figure}\centering
\includegraphics[width=0.4\textwidth]{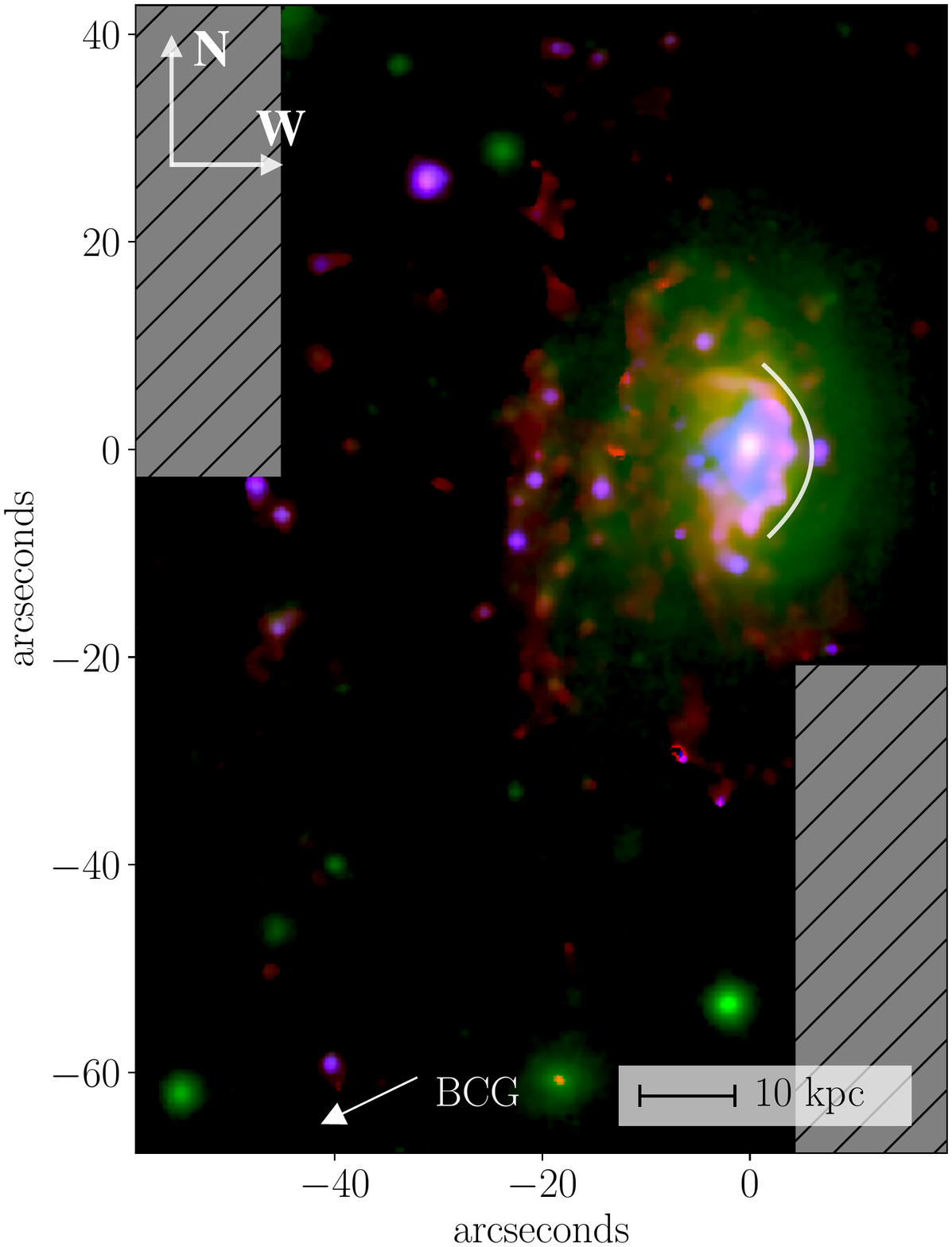}
\caption{Composite image of JO201, red: $\mathrm{H}\alpha$, green: stellar continuum fitted between 80-200\AA{} either side of the H$\alpha$ line, blue: [OIII]. The emission lines have been masked to a S/N ratio of 3 or greater and cleaned using a $5 \times 5$ median filter. The white curve indicates the leading edge of the galaxy, where the gas is compressed by RPS. The large offset of the red H$\alpha$ from the green stellar component is clearly visible, as well as the large concentration of [OIII] in the central 2-3kpc of the disk. The bright [OIII] emission in the central region of the disk corresponds with a drop in the SFR according to both the H$\alpha$ luminosity shown in the figure and the stellar age shown later. This may be due to feedback from the AGN inhibiting star formation in the central region of the disk}\label{figure:ha_stellar_oiii}
\end{figure}

\subsection{The jellyfish galaxy JO201} \label{subsec:jo201}


JO201 is a massive spiral galaxy located in the massive galaxy cluster A85 ($\mathrm{M}_{200}=1.58\times10^{15}$, velocity dispersion $\sigma_{\rm cl}=982\mathrm{km~s}^{-1}$) at a redshift $z_{cl}=0.05586$. 
The galaxy has a stellar mass of $~3.55\times 10^{10}\mathrm{M_\odot}$ and a dynamical mass of $~5.10\times 10^{11}\mathrm{M_\odot}$ (GASP II).

As also shown in GASP II, JO201 lies at a projected distance of 360~kpc from the cluster centre, and has a line-of-sight velocity of 3363.7km~s$^{-1}$ (or $3.4\times \sigma_{\rm cl}$) with respect to the mean velocity of the cluster. The proximity to the cluster centre, coupled with its extreme velocity place the galaxy close to the peak of stripping \citep[see][]{Jaffe2018}. Moreover, the inclination of the galaxy and the morphology of its tails indicate that the interaction with the ICM is close to face-on.
The MUSE data further reveals  projected tails of $\mathrm{H}\alpha$ emitting gas up to at least $\sim 50~\mathrm{kpc}$ in length (see Figure~\ref{figure:ha_stellar_oiii}), structured into kinematically cold knots (velocity dispersion $<40~\mathrm{km~s}^{-1}$) and warm diffuse emission ($>100~\mathrm{km~s}^{-1}$). These tails rotate with the stellar disk out to $\sim 6~\mathrm{kpc}$, but become increasingly redshifted in the disk outskirts (by up to $900~\mathrm{km~s}^{-1}$) with respect to the centre of the galaxy due to the intense ram-pressure along the line of sight, that causes the gas to drag behind the galaxy.

It is likely that JO201 is falling into the cluster within a small group of galaxies (GASP II) and that its visible nuclear activity (shown in Section~\ref{sec:analysis}) is being triggered by the interaction with the ICM \citep{Poggianti2017b}.

Follow-up studies of JO201 at different wavelengths have further revealed the presence of extra-planar molecular CO gas \citep{Moretti2018} and in-situ star formation activity visible in the Ultra Violet \citep[UV;][]{George2018}. 
Moreover, \citet{Venkatapathy2017}, 
detected 
emission in the tails of JO201 (A85[DFL98]176) at blue and UV wavelengths, but not in near infrared, which suggests a lack of an older stellar component, as expected for hydrodynamic disruption.

Fig~\ref{figure:ha_stellar_oiii} shows an overview of JO201 with the stellar continuum (fitted between 80 and 200\AA{} from the H$\alpha$) in green, H$\alpha$ in red and [OIII] in blue, illustrating the intense stripping by the large offset of the H$\alpha$ component from the stellar component, and the intense [OIII] emission from the central AGN.

\subsection{MUSE datacube}\label{sec:MUSE}

JO201 was observed by MUSE as part of the GASP programme on December 17, 2015 with photometric conditions.
A total of eight exposures of 675s each were taken across two adjacent fields to produce the final mosaic. The data were reduced using ESOREX recipes version 3.12, MUSE pipeline version 1.2.1 (see GASP I for details on the reduction process). The sky background was measured from the sky coverage within the individual frames for subtraction and the final wavelength and flux calibrated images were combined using sources in the white light images for alignment. The final image quality of the reduced cube is 0."8 FWHM.




To prepare the cube for emission line fitting, the effect of dust extinction was accounted for accordingly with other GASP studies. The full details are outlined in GASP I. Dust extinction by the Milky Way was corrected across the cube, and internal extinction within JO201 was corrected over individual spaxels.

The resulting correction was applied only to regions in which all corrected lines are observed to S/N $>3$ and, since the correction only applies to regions which are dominated by photoionisation, spaxels which lie above the \citep{2001ApJ...556..121K} line on the [OIII]/H$\beta$ vs [NII]/H$\alpha$ \citep*[][BPT]{Baldwin1981} diagram were also omitted.

Much of the attenuation by dust (map omitted) occurs within the regions of the disk and it is slightly less prominent in the trailing knots. This pattern is similarly observed in other GASP galaxies undergoing stripping along the plane of the sky.



\begin{figure*}\centering
\includegraphics[width=\textwidth]{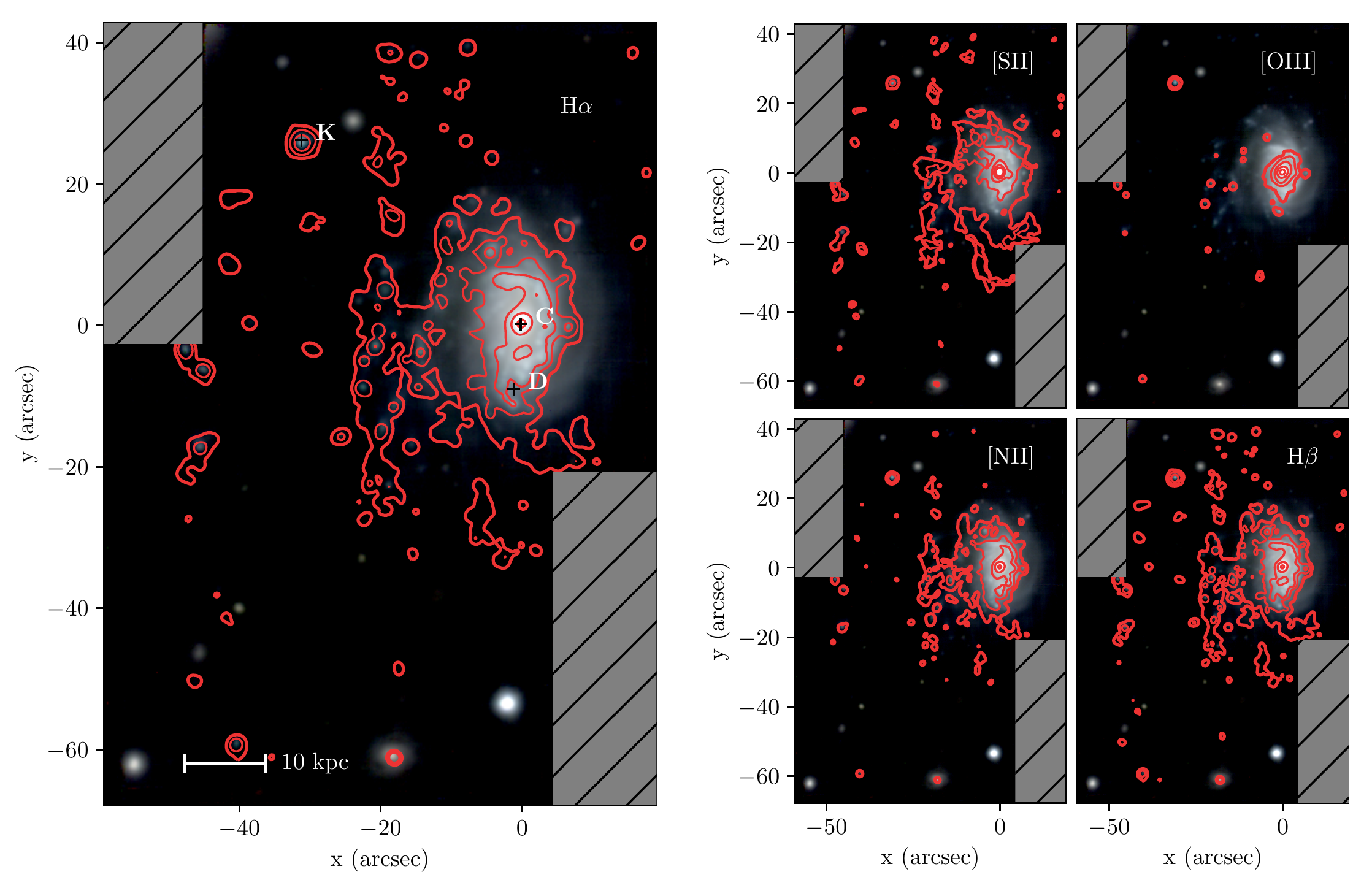}
\includegraphics[width=\textwidth]{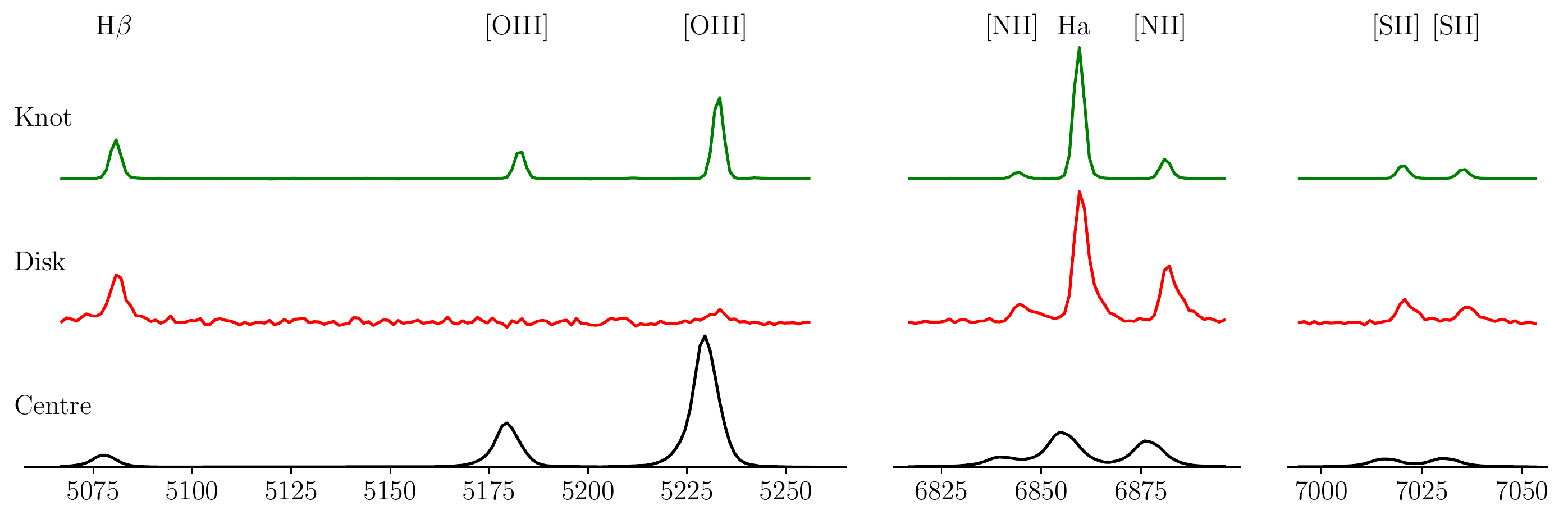}
\caption{\textit{Top:} Emission-line  maps of H$\alpha$, [SII], [OIII], [NII] and H$\beta$ from the dust-corrected emission-only cube are shown in red contours, on top of an RGB image of JO201 produced using $1000\AA$ slices integrated from the datacube (B:5000-6000\AA, G:6000-7000\AA, R:7000-8000\AA). Each line has been masked to a S/N of 3 or greater. 
\textit{Bottom:} Examples of spectra, with stellar component subtracted as described in the text. The spectra correspond to 3 regions of interest indicated on the H$\alpha$ linemap in the left panel: Galaxy centre (C), disk (D), and a knot (K). 
The central region exhibits much broader lines around the AGN, with a larger proportion of [OIII] emission. Within the disk, the lines are much narrower and slight asymmetry is visible due to the stripping acting in the line of sight. The knot exhibits narrow emission with strong H$\alpha$ and [OIII] emission.
}
\label{figure:linemaps}
\end{figure*}

\subsection{Emission-only datacube}\label{sec:stellarabs}
The emission line fluxes are subject to stellar absorption and direct measures will underestimate the true flux, unless the absorption line contribution can be accurately estimated.
To obtain emission line fluxes corrected for stellar absorption, we ran the spectrophotometric code {\sc sinopsis} \citep{Fritz2017}, which includes the pertinent nebular emission lines. {\sc sinopsis} was used to find the best fitting combination of single stellar population (SSP) models to the spectra.

The best-fitting solution found by {\sc sinopsis} was then subtracted from the original cube to produce an emission-line-only cube which was used in the subsequent analysis. A few examples of the continuum subtracted spectra from this cube are shown in the bottom panel of figure \ref{figure:linemaps}.



\section{Emission-line fits}\label{sec:elfits}



As in GASP II, we used the IDL custom code \textsc{kubeviz} \citep{2016MNRAS.455.2028F} written by M. Fossati and D. Wilman to fit the emission lines, adding some changes to optimise for the line ratio analysis. \textsc{kubeviz} fits selected emission lines to the spectrum of each spaxel with models comprised of one or two Gaussian profiles, or by defining lines using moments. The full details of the fitting process are described in GASP II. In summary, we spatially smoothed the emission-only datacube (c.f.~Sec.~\ref{sec:stellarabs}) using the mean value within a $3 \times 3$ kernel. 
The zero-point redshift 
was set to $z=0.045$, the mean redshift for JO201, and the option in \textsc{kubeviz} was enabled which scales the noisecube errors to give a reduced $\chi^2=1$. 

It is important to mention that the large line-of-sight component of the stripping in JO201 causes the emission-lines in some regions of the disk to be non-Gaussian (see ``Disk" spectrum in Figure 3, and also GASP II Figure 5), often requiring two-component fits. Typically, in these regions the emission is described by a sharp narrow line accompanied by a fainter component with a higher velocity dispersion and a slightly redshifted line-of-sight velocity.

As shown in GASP II, most of the spaxels requiring double-component fits are located around the edges of the disk, where ram-pressure has already removed and spread a significant amount of gas. The broad emission seen in the centre of the galaxy on the other hand originates from the active galactic nucleus (AGN).

Both single and double-component fits were carried out to the emission lines present in the MUSE spectra.
The fits used in the subsequent analysis were therefore a combination of single and double component fits, decided on a spaxel-by-spaxel basis depending on the line profiles. 
In any given spaxel, the double-component fits were preferred if the two components were detected to S/N $> 3$ and they are sufficiently separated in velocity (i.e. detectable emission-line wings, or multiple peaks in the same line of sight). Otherwise, the single-component fits with S/N $> 3$ were used. The selection process between 1 and 2 component fits is described in further detail in GASP II. 
It is worth noting that in the regions where the emission is faint, the single-component fits tend to yield a higher signal-noise ratio.

For the rest of the analysis, the two components of the emission-line fits will be referred to as ``primary" and ``secondary". The primary component refers to either the single component fit or the narrower component (velocity dispersion: 3.0 - 310.7km~s$^{-1}$, mean value: 84.4km~s$^{-1}$) of the double component fit. These together describe the majority of the emission from the galaxy disk, knots and tails. The secondary component instead, refers to the more diffuse component (velocity dispersion: 21.4 - 376.8km~s$^{-1}$, mean value: 123.4km~s$^{-1}$) of the two component fit and generally describes redshifted tails caused by line-of-sight stripping, co-spatial knots, or emission around the AGN region. Since both components have a range of velocity dispersion, and often have comparable widths (but are simply shifted in velocity), we have avoided referring to them as ``narrow" and ``broad".


In the kinematic study of JO201 (GASP II) we focused on the $\mathrm{H}\alpha$ component fitted alongside the N[II] lines. In this study we fit all lines shown in Table \ref{table:emission_lines} in order to construct line ratio diagnostics, to obtain the metallicity and ionisation parameter.

For most of the analysis, since all lines were required for each diagnostic, the cube was masked to include only areas in which all required emission lines were detected in \textsc{kubeviz} to a signal-noise ratio of 3 or above. This was in addition to the internal flag system used by \textsc{kubeviz} which excludes fits with zero velocity errors or velocity dispersions narrower than the instrumental resolution.

The resulting line flux maps for H$\alpha$, [SII], [OIII], [NII] and H$\beta$ are shown at the top of Figure \ref{figure:linemaps}. 
The H$\alpha$ map prominently highlights the star forming knots and steep leading edge on the disk, whereas the [OIII] line shows the strong emission originating from the AGN at the galaxy centre.

\begin{table}
\centering
\caption{Emission Lines considered in this analysis. Table columns are: (1) Name of emission line, with ``forbidden" transitions denoted by square brackets. (2) Wavelength in air}
\begin{tabular}{l c}\label{table:emission_lines}
\\\hline
Line & $\lambda$ (\AA)\\
\hline
H$\beta$ &4861.33\\
{[}OIII{]} &4958.91\\
{[}OIII{]} &5006.84\\
{[}OI{]} &6300.30\\
{[}OI{]} &6363.78\\\hline
\end{tabular}
\begin{tabular}{l c}
\\\hline
Line & $\lambda$ (\AA)\\
\hline
{[}NII{]} &6548.05\\
H$\alpha$ &6562.82\\
{[}NII{]} &6583.45\\
{[}SII{]} &6716.44\\
{[}SII{]} &6730.81\\\hline
\end{tabular}
\end{table}


\begin{figure*}\centering
\includegraphics[width=0.88\textwidth]
{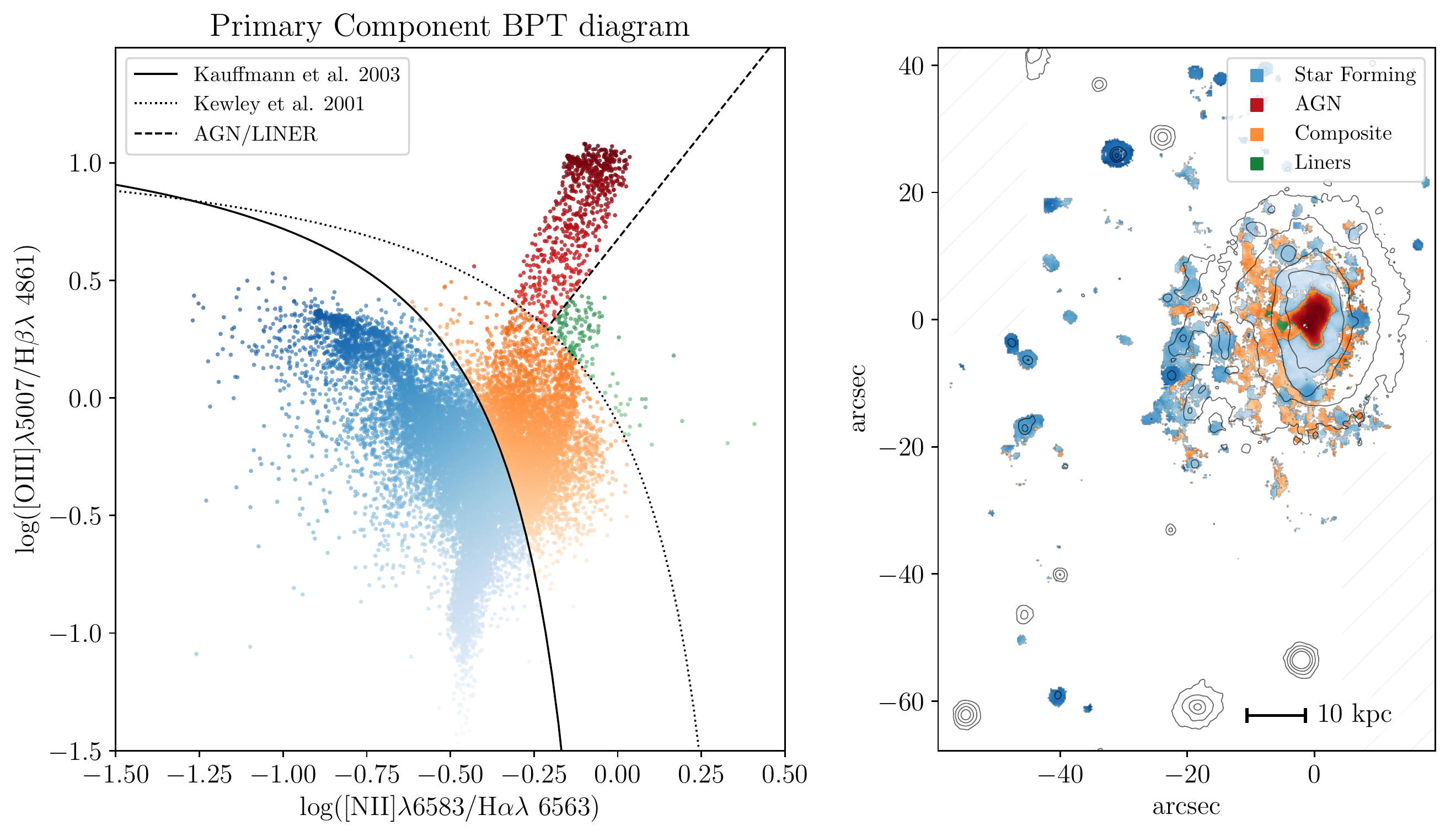}
\includegraphics[width=0.88\textwidth]{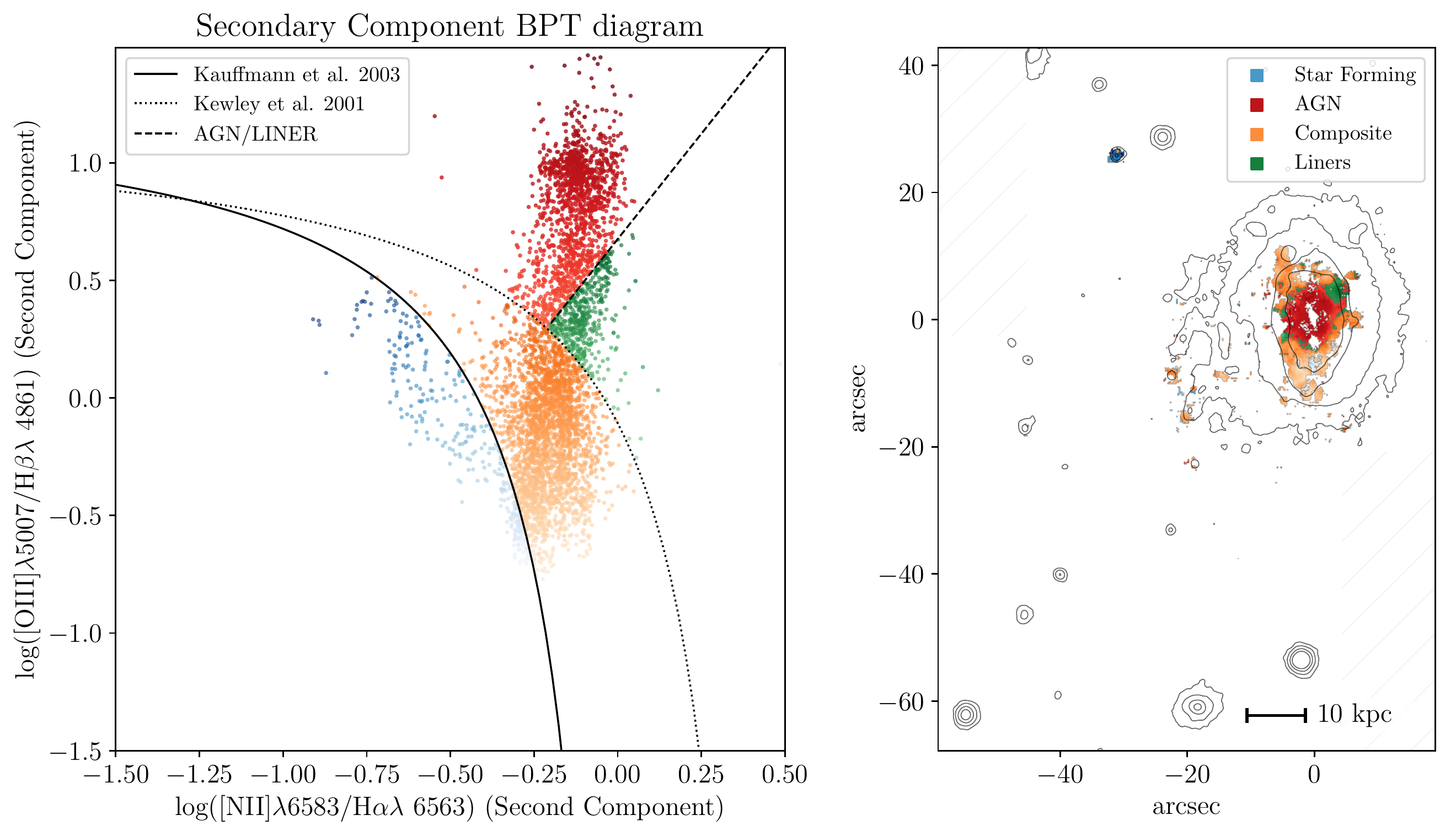}
\caption{[OIII], [NII] BPT line-ratio diagnostics for all spaxels with S/N $>3$ on the primary (top) and secondary (bottom) component of the two-component fits (left), mapped on top of the galaxy (right). The black lines separating different ionisation sources on the left panel come from: \citet{Kauffmann2003}, to separate star formation dominated regions from composite; \citet{Kewley2001}, to separate composite from AGN/LINER regions; and the AGN/LINER separator, taken from \citet{SharpBlandHawthorn2010}. The contours on the right panel correspond to the stellar continuum fitted around the $\mathrm{H}\alpha$ line for context. The colour gradients are included to show variations within each region of the BPT diagram. The distribution of the ionised gas in the galaxy prominently shows an AGN region in centre of the galaxy, and knots with  ongoing star formation, that are surrounded by ionised gas lying in the composite region of the BPT diagram.
}\label{figure:BPT_OIII}
\end{figure*}


\begin{figure*}\centering
\includegraphics[width=\textwidth]
{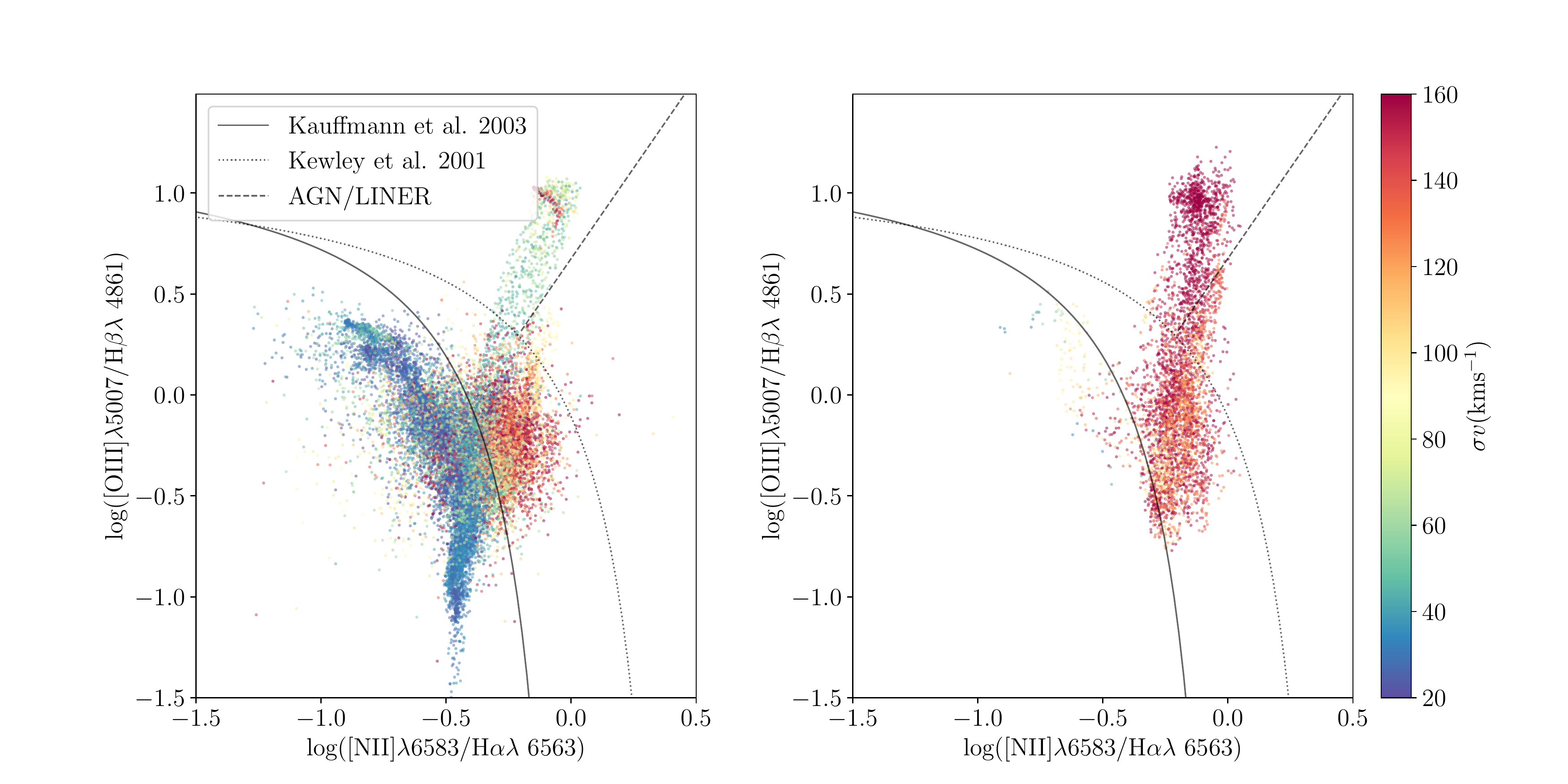}
\caption{[OIII], [NII] BPT line ratio diagnostics for the primary (left) and secondary (right) component for all spaxels with primary component detected to S/N $< 3$, coloured by $\mathrm{H}\alpha$ velocity dispersion. The dark blue clumps on the star forming branch in the left panel indicate the denser star forming knots in the galaxy's tail, while the large cluster of points with velocity dispersion 120 to 150 km~s$^{-1}$ in the composite region is comprised of ionised gas resulting from the interaction with the ICM. Finally, the broader lines are naturally located in the AGN-ionised region.
In the right panel the secondary component, which traces the stripped disk and the nuclear region, is located between the composite region, consistent with turbulent, excited gas being stripped from the galaxy, and the AGN region to the upper right of the figure.
}\label{figure:bptsigma} 
\end{figure*}

\section{Emission-line Diagnostics}\label{sec:analysis}
\subsection{Gas ionisation mechanism}

BPT
line diagnostic diagrams were produced from the $\mathrm{H}\alpha$, {[}OIII{]},
 {[}NII{]} and $\mathrm{H}\beta$ lines to distinguish the 
ionisation mechanism of the nebular gas.

Fig~\ref{figure:BPT_OIII} shows the $[\mathrm{OIII}]/\mathrm{H}\beta$, $[\mathrm{NII}]/\mathrm{H}\alpha$ line diagnostic diagram for the primary (top) and secondary (bottom) components of the two-component fits. The left side of the figure shows the position of each spaxel in BPT space, coloured by the excitation mechanism according to the intersecting lines on the figure. The star forming/composite/AGN regions are defined by the \citet{Kauffmann2003} and \citet{Kewley2001} lines and the AGN/LINER separator is taken from \citet{SharpBlandHawthorn2010}. The plot on the right shows these spaxels in situ on the galaxy, revealing the regions in which each ionisation mechanism dominates.

The majority of the AGN emission comes from the very central region of the galaxy disk. The composite emission emanates from the diffuse gas which surrounds the disk and between the knots, which themselves are dominated by star formation. A small cluster of spaxels dominated by LINER type sources is visible on the galaxy to the left of the AGN dominated central region.
%
Most of the secondary component lies within the composite region. This generally corresponds to the diffuse gas surrounding the disk and knots as seen in the primary component, but also comes from broad velocity tails in the line of sight coming from the edges of the galaxy disk, where the line-of-sight stripping produces a tail in the velocity distribution (see GASP II).

To illustrate this, we have shown in Figure \ref{figure:bptsigma} the same BPT maps as in the left panels of Figure \ref{figure:BPT_OIII}, this time coloured by the velocity dispersion of the gas, as measured from the broadness of the emission-line fits in GASP II. In the case of the primary component this illustrates that the knots are kinematically colder (see also Fig. 13 of GASP II), consistent with the idea that the knots condense out of the stripped gas, collapsing and forming stars, as confirmed by UV observations which show in-situ star-formation in the stripped tails \citep{George2018}. These kinematically cold, star forming knots are a feature of other GASP jellyfish galaxies (Poggianti et al. 2018 in press) and other strongly ram-pressure stripped Jellyfish galaxies in the literature such as ESO 137-001 \citep{Fumagalli2014,2016MNRAS.455.2028F}, UGC6697 \citep[stripped edge-on;][]{Consolandi2017} and SOS 114372 \citep[][also included in the GASP sample as JO147]{Merluzzi2013}.

The AGN region exhibits broader lines in general and this peaks within the cluster of AGN-dominated points at the upper right. The cloud of higher velocity dispersion points within the composite region results from the diffuse stripped gas which surrounds the disk and the knots. In the case of the secondary component, which traces mainly the diffuse or nuclear gas, the majority of points lie within the composite and AGN dominated regions, originating from the stripped gas in the outer disk, and the broad AGN emission in the centre respectively. 

\subsection{Metallicity and Ionisation Parameter}

Using the software pyqz \citep{Dopita2013} the metallicity (12 + log(O/H)) and ionisation parameter (log Q) were calculated for each spaxel using the direct estimate interpolated from the [NII]/[SII]+ vs [OIII]/[SII]+ diagnostic (where [SII]+ refers to the sum of the two SII components). The metallicity and ionisation parameter measurements are only valid in HII regions, therefore spaxels outside the star forming region of the BPT diagram were masked out. The pyqz python module uses a finite set of diagnostic line ratio grids produced using the MAPPINGS code, onto which the selected combination of line ratios are interpolated. We used a version of pyqz modified to use the MAPPINGS IV grids which are calibrated over a larger range $7.39 < 12+\log(\mathrm{O}/\mathrm{H}) < 9.39$ than the MAPPINGS V grids (F. Vogt, priv comm.).

\begin{figure}\centering
\includegraphics[width=0.5\textwidth]{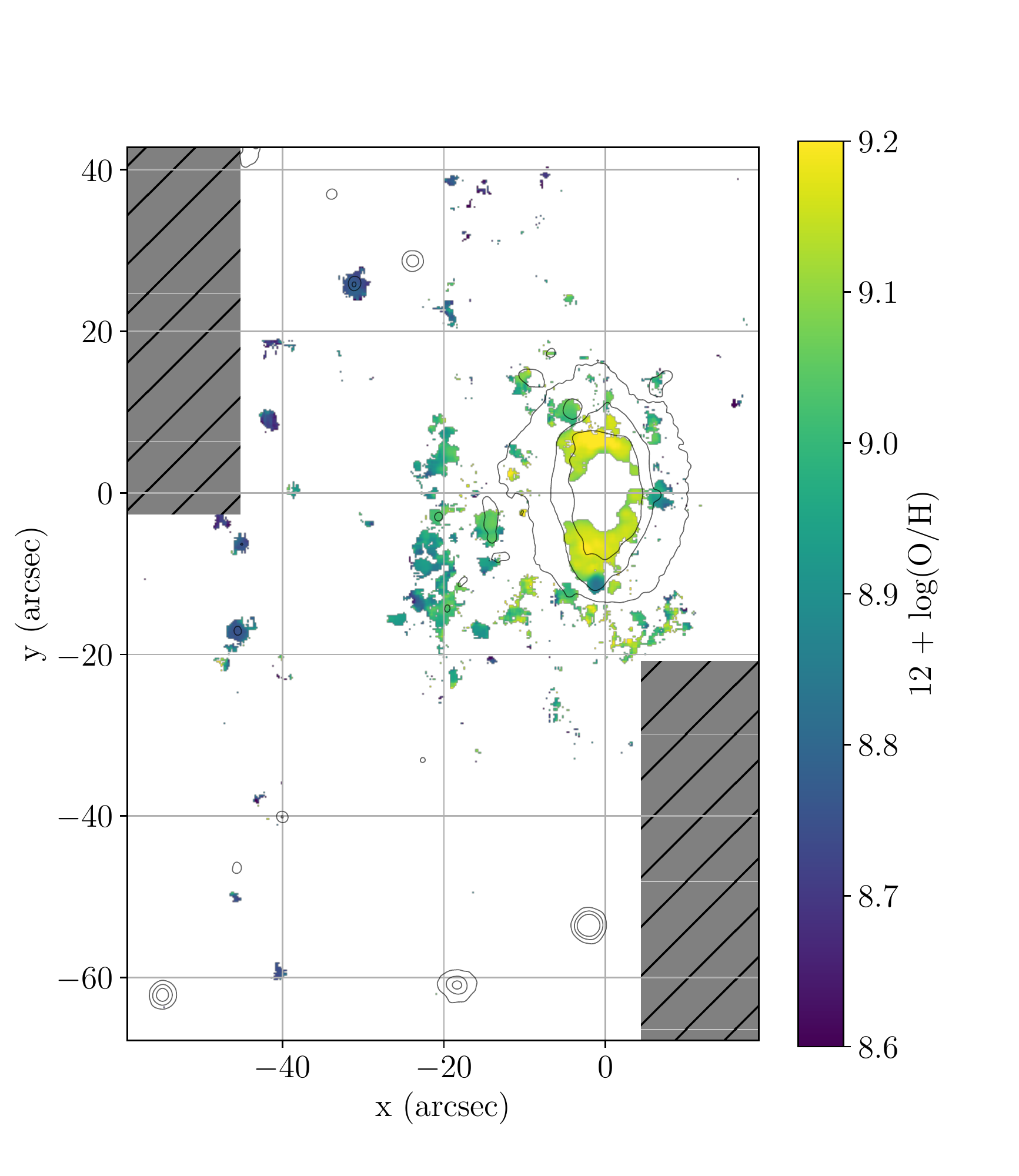}
\caption{Metallicity
of stripped gas for HII regions calculated using pyqz as described in the text. Black contours show stellar continuum fitted around the H$\alpha$ line for context. The highest metallicities are visible in the disk of the galaxy, with lower metallicities in the tails and knots. This outwardly-decreasing trend is further explored in Fig~\ref{figure:z_radius} for the star-forming knots. 
}\label{figure:logqpyqz}
\end{figure}


The results from pyqz for the metallicity are shown in Figure~\ref{figure:logqpyqz}.
We find that the metallicity varies by $\sim 0.6$ dex in the ionised gas and has similar values to those in other GASP jellyfish galaxies, such as JO204 \citep{Gullieuszik2017} and JO206 \citep{Poggianti2017}. The most metal-rich gas is found in the disk of the galaxy up to values of $\sim 9.2$, with intermediate metallicities of $\sim 8.9$ in the stripped gas. The most metal-poor gas resides in the more distant knots $> 40$kpc from the galaxy, with $12+\log(\mathrm{O}/\mathrm{H}) \approx 8.7$. For comparison, the \citet{Tremonti2004} relation gives an average metallicity of 9.06 for a galaxy of this stellar mass. The outwardly-decreasing trend is better seen in the star-forming knots (Section~\ref{sec:knots}). Comparing with undisturbed galaxies, JO201 has a higher central metallicity compared to galaxies observed with CALIFA in \citet{Sanchez-Menguiano2017}, (with stellar masses ranging over 9.18-11.30M$_\odot$) whilst the range of metallicity values within JO201's disk and tails is similar to that of the arm regions of the CALIFA galaxies.


The ionisation parameter (not shown) is low in general with a median value of $(\sim 7.1)$ in the disk and of $(\sim 7.3)$ in the tails. These values, like the metallicity, are similar to those observed in other strongly stripped GASP jellyfish. 


\begin{figure*}\centering
\includegraphics[width=\textwidth]{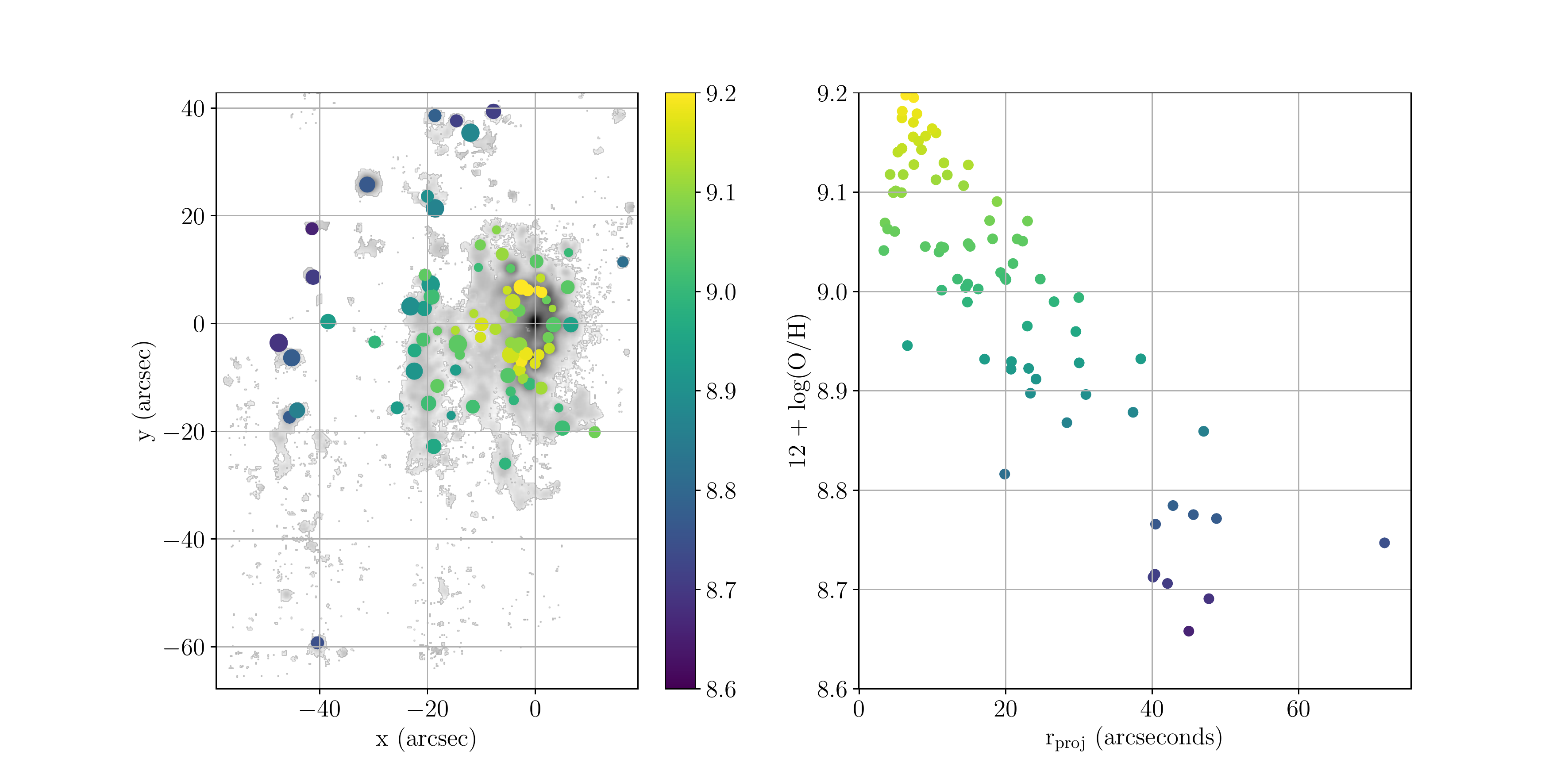}
\caption{\textit{Left:} Metallicity calculated using pyqz direct estimate for integrated spectra of SF knots shown over grey $\mathrm{H}\alpha$ linemap. \textit{Right:} Metallicity of knots from left panel against their projected distance $\mathrm{r}_{\rm proj}$ from centre of disk. The figure shows the metallicity of the knots decreasing outwardly from the galaxy centre, with more distant knots having generally lower metallicities.
}\label{figure:z_radius}
\end{figure*}

\section{Star Forming Knots}\label{sec:knots}

The bright star forming knots within the tail of the galaxy are a common feature in many GASP sample galaxies and other observed jellyfish galaxies. To further study the emission properties of these knots, the spectra within circular masks were binned spatially and lines were fitted to the resulting total spectrum using the \textsc{kubeviz} mask function, which fits emission lines to the integrated spectra within each mask. The positions and radii of each knot were obtained using a custom script, described in full detail in \citet{Poggianti2017}, in order to produce the masks. The emission line fluxes measured by \textsc{kubeviz} were subsequently corrected for internal dust extinction using the same calculation as described in section \ref{sec:MUSE}. In total, 85 knots were found, of which 63 lie within the star-forming branch below the Kauffmann line marked on the line ratio diagram. The measured line fluxes were fed into pyqz to calculate direct estimates of the metallicities.

The left side of the plot in Figure~\ref{figure:z_radius} shows the spatial position of each knot against the galaxy. The size of each point scales with the size of each knot and the colour is indicative of the direct estimate of the metallicity as computed by pyqz.
The plot reveals a general decrease in the metallicity of the knots with distance from the galaxy disk, with the highest metallicities in the disk and lower values in the most distant knots. This outwardly decreasing trend is better visualised in the right plot of the figure, which shows the metallicity of each knot against the projected distance from the galaxy centre. 
The metallicity gradient revealed by the knots in JO201 
is similarly observed in other GASP galaxies 
(see Franchetto et al. in prep.), and it is
consistent with the mechanism of outside-inward stripping, whereby the most distant, lowest metallicity gas from the galaxy is stripped first, followed by subsequently higher metallicity gas closer to the centre. 
In fact, semianalytic models  of galaxies in clusters \citep{Gupta2017} have invoked outside-in gas stripping to explain, at least partially,  
the cluster scale (integrated) metallicity gradients observed in galaxy clusters \citep[see e.g.][]{Ellison2009}.

For reference, the distribution of the star-formation rate, gas density, and gas mass for the knots in JO201 are shown and discussed in the Appendix. Overall we find similar values to those found in other heavily stripped jellyfish galaxies from GASP.

\begin{figure}\centering
\includegraphics[width=0.5\textwidth]{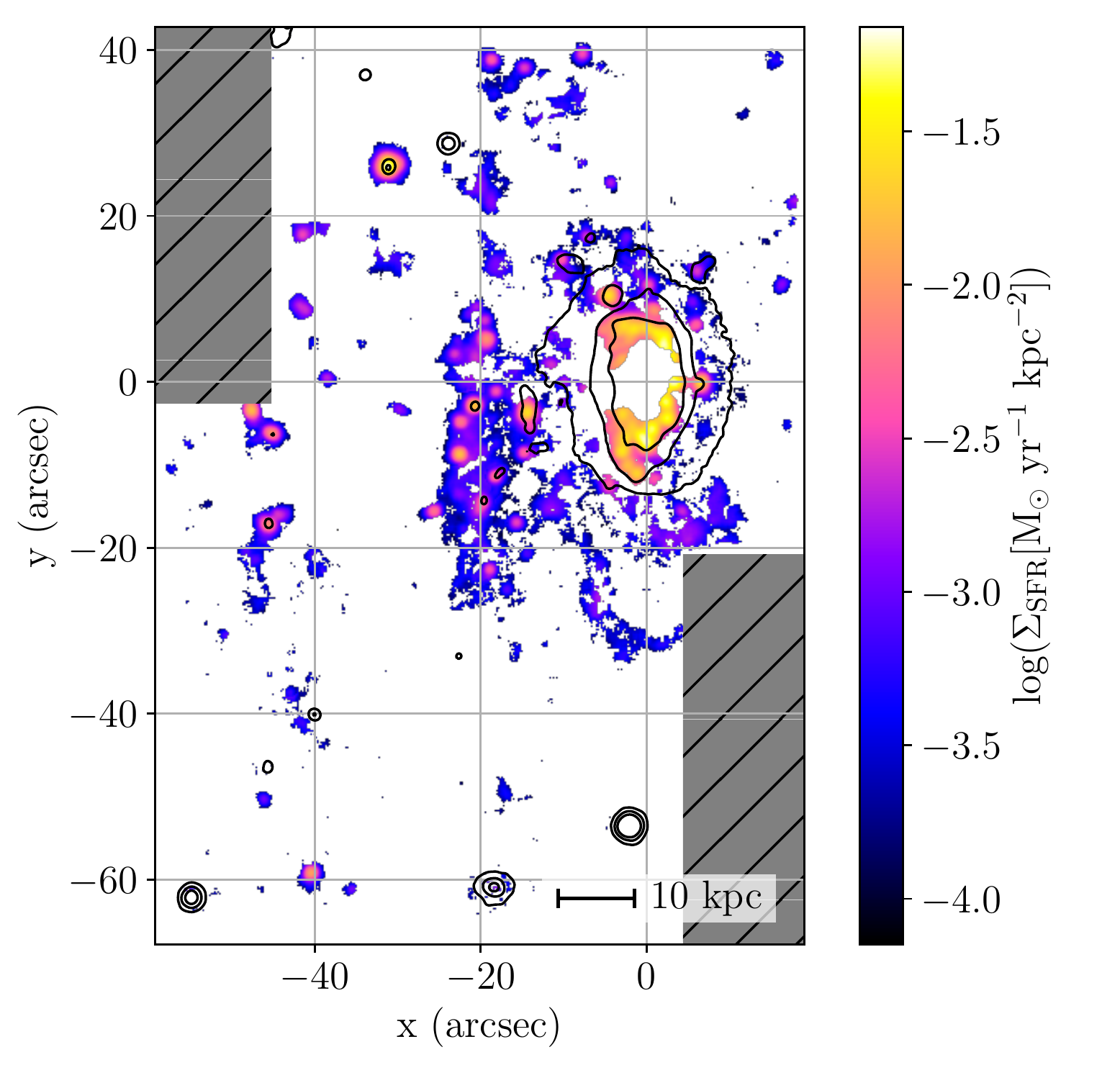}
\caption{Star-formation rate surface density within JO201 calculated for each spaxel
($\mathrm{M}_\odot \mathrm{yr}^{-1} \mathrm{kpc}^{-2}$).
Black contours show stellar continuum fitted around the H$\alpha$ line for context.
We observe high star-formation rates in the leading (western) edge of the disk, triggered by the interaction with the ICM. High SFRs are also seen in the collapsed knots trailing behind the galaxy and in particular within the bright knot in the north-eastern region of the image.}\label{figure:sfr}
\end{figure}

\begin{figure}\centering
\includegraphics[width=0.5\textwidth]{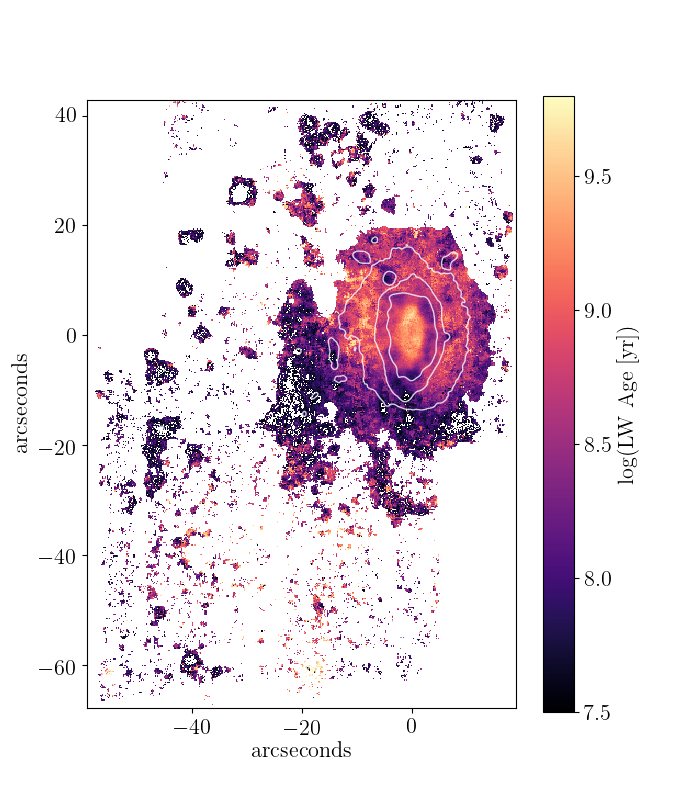}
\caption{Luminosity weighted stellar age (colourbar) shown along with of stellar continuum isophotes (white contours). A clear age gradient is seen, with the oldest stars residing around the disk and lower ages in the tails, whilst a ring of young stars is visible within the disk surrounding the centre.}\label{figure:lwage}
\end{figure}

\begin{figure*}\centering
\includegraphics[width=0.96\textwidth]{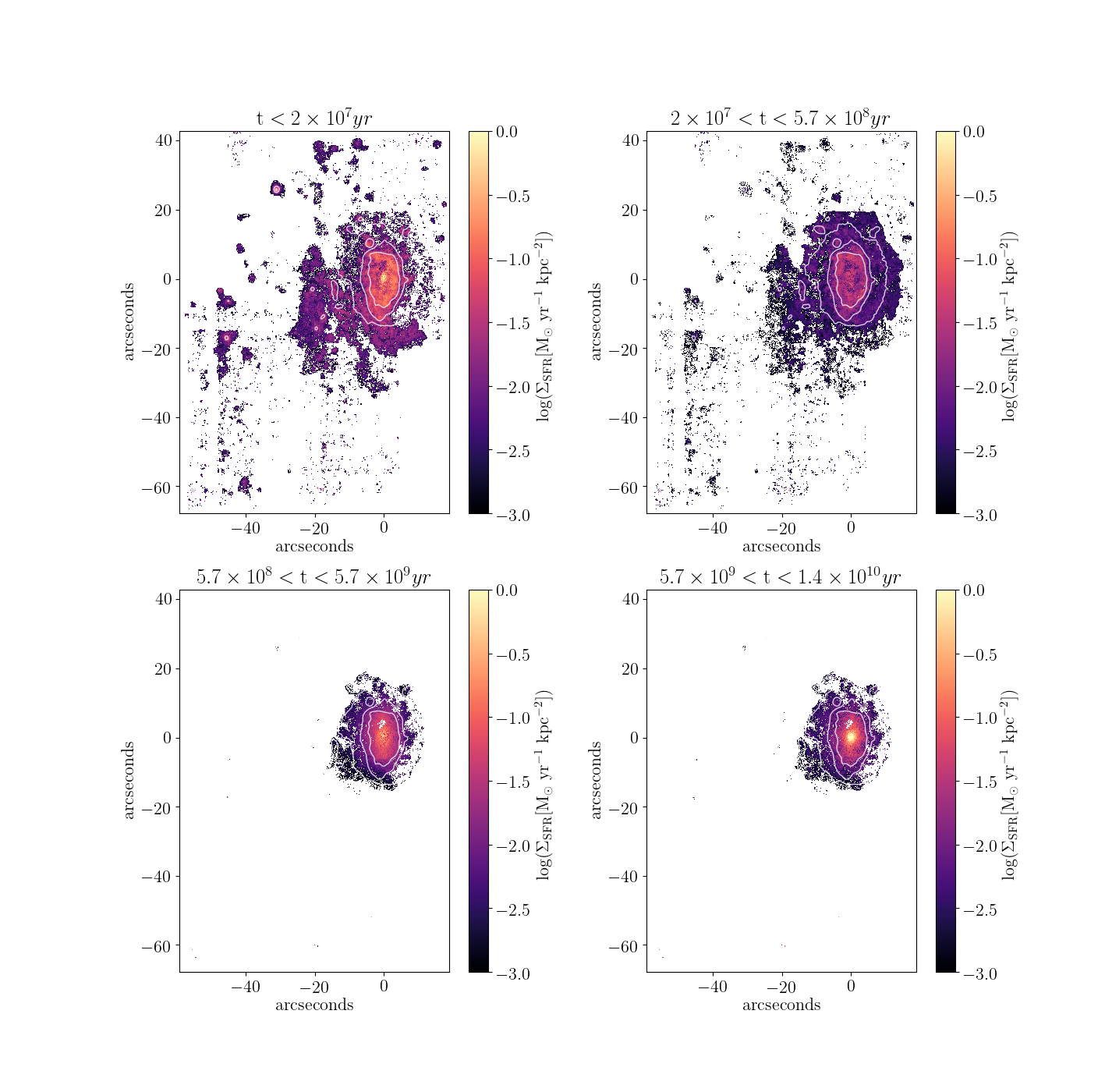}
\caption{Star-formation rate surface density grouped into several stellar age ranges, indicating the average SFR per kpc$^2$ during the last $2 \times 10^7 yr$ (top left), between $2 \times 10^7 yr$ and $5.7 \times 10^8 yr$ (top right), $5.7 \times 10^8 yr$ and $5.7 \times 10^9 yr$ and $> 5.7 \times 10^9 yr$ ago. The top two plots have been masked to S/N$>0.5$, whilst the bottom plots have been masked to spaxels of S/N$>3$ and with a contribution to the stellar continuum above 5\% The figure reveals recent star formation occurring in the tails and knots, as well as the leading edge of the disk, with the oldest stars residing in the bulge and disk of the galaxy. White contours indicate continuum isophotes.}\label{figure:sfrd} 
\end{figure*}


\section{Star formation}\label{sec:sf}

It is interesting to examine the distribution of star-formation activity within the galaxy, since it can reveal information about the ram-pressure stripping history and the disruption and compression of the gas component.

The star-formation rates for each spaxel
are shown in figure \ref{figure:sfr}.
The star-formation rate surface density, $\Sigma_{\rm SFR}$, is calculated using the H$\alpha$ luminosity, corrected for dust extinction and stellar absorption, in the \citep{1998ARA&A..36..189K} method transformed to a Chabrier IMF, for each individual spaxel with an H$\alpha$ S/N $>$ 3.

The SFR surface density values across JO201 are typically around $0.001\mathrm{M}_\odot \mathrm{yr}^{-1} \mathrm{kpc}^{-2}$ for diffuse regions and $0.02\mathrm{M}_\odot \mathrm{yr}^{-1} \mathrm{kpc}^{-2}$ for denser blobs and regions within the galaxy disk. The bright trailing knots to the east of the galaxy are therefore strongly star forming, with a particularly active region around $5\mathrm{kpc}$ in diameter to the north east of JO201. As well as within the trailing knots, a ring of strong star formation is apparent within the disk, where the gas is likely to have been compressed by interaction with the ICM. Within this ring a decrease in the star formation is coincident with the AGN emission, suggesting AGN feedback suppressing star formation in this region. This is also suggested by the deficit of $\mathrm{H}\alpha$ emission coincident with the [OIII] emission in Figure \ref{figure:ha_stellar_oiii} and will be further investigated in George et al.~(in preparation).

Overall, the range and spatial distribution of the star-formation rates obtained for JO201 are comparable with those observed in strongly stripped galaxies which are moving on the plane of the sky such as JO206 \citep{Poggianti2017} and JO204 \citep{Gullieuszik2017} in the GASP sample. In particular, the dense regions of intense star formation are often observed in the tails of jellyfish galaxies. \citep{Fumagalli2011,Kenney2014,Boselli2018}

\citet{Vulcani2018} found that the total integrated SFR within the galaxy and trailing gas 
is $6\pm 1\mathrm{M}_\odot \mathrm{yr}^{-1}$, which is high for the stellar mass of the galaxy. In fact, the same authors find that the SFR in the disk of JO201 is around 0.4-0.5 dex above the main sequence of a control sample of non-stripped disk galaxies, which is among the highest offsets found in the sample. 

Moreover, \citet{Moretti2018b} presented APEX observations of JO201, finding CO emission in the disk that partly traces the velocity of the stellar disk, 
and partly the velocity of the stripped H$\alpha$ knots. 
Further from the disk the molecular gas does not follow the velocity of the H$\alpha$ gas. The large depletion times measured for the molecular gas suggest that the majority will be dispersed into the ICM.

In comparison with the far ultraviolet observations presented in \citet{George2018}, which have lower spatial resolution than the MUSE data presented in this paper, we find that although the measured SFR values differ by 0.1-0.5M$_\odot \mathrm{yr}^{-1}$, the two methods reveal the same trends and the same pattern of star formation localised in collapsed knots. The extinction in UV and Halpha follows different laws. Since the UV extinction was derived using the balmer decrement, there may be uncertainty in the correction, accounting for the deviation in the SFR values.

The bright knots in the tails of JO201, which have high velocities relative to the stellar disk, therefore appear to 
have originated from in-situ star formation happening in the stripped (trailing) gas. 

\section{Stellar Populations}\label{sec:stellar}

We calculated the ages of the stellar populations across the galaxy using the {\sc sinopsis} spectrophotometric fitting code \citep[c.f. Section \ref{sec:stellarabs}, and Paper III:][]{Fritz2017}.
A map of the luminosity weighted stellar age in JO201 is shown in Figure~\ref{figure:lwage}. The plot shows a ring of lower stellar age approximately 3-5kpc in radius around the centre of the disk and a decreasing stellar age into the tails compared with the outer edges of the disk. The decrease in stellar age along the leading edge and in a ring around the disk is consistent with a burst of recent star formation occurring due to compression of the gas as the galaxy collides with the ICM. This increases the fraction of young stars in that region and lowers the observed luminosity weighted age.

Beyond the outermost contours of the stellar continuum, the older stellar population rapidly diminishes and the tails contain only young stars. The stellar age further decreases in the collapsed knots within the stripped gas, as a result of the onset of star formation. This gradient in stellar age from the disk to the tails is similarly observed in JO206 \citep{Poggianti2017} and JO204 \citep{Gullieuszik2017}.

From the {\sc sinopsis} fits, we further computed the star-formation rate surface density in different bins of stellar age with the intention of seeing the consequences of ram-pressure stripping on the star-formation history of the galaxy. This is shown in Figure \ref{figure:sfrd}.  The two oldest stellar component bins (between $5.7 \times 10^8 yr$ and $5.7 \times 10^9 yr$ ago, and $> 5.7 \times 10^9 yr$) are dominated by star formation in the central region of the galaxy and no stellar component of this age is detected outside the disk (i.e. the tails are not seen). 
More recent bursts of star formation (between $2 \times 10^7 yr$ and $5.7 \times 10^8 yr$ ago) can be seen along the leading edge of the disk and some of the stripped tails. Such activity is very likely triggered by ram-pressure as the galaxy first interacted with the dense ICM. Subsequently, in the last $2 \times 10^7 yr$, star-formation appeared in the knots within the long tails of stripped gas. 

Our findings are similar to the distribution of stellar populations seen in \citet{Poggianti2017} within JO206, with older stellar populations dominating the disk and a younger stellar component reaching further into the tails. A lack of very young stars can be seen along the leading edge of the galaxy, followed by a region where the gas has been compressed or stripped by the ICM. This can also be seen in JO204 \citep{Gullieuszik2017}.


\section{Stripping Timescale and length of the stripped tails}\label{timescale}


It is possible to use different observable parameters of the galaxy and its host cluster, to estimate the time that the galaxy has been experiencing gas stripping. In this section we compare several independent calculations. 

\subsection{Stripping timescale from an orbital analysis}
\label{sec:timescale1}

In previous GASP studies \citep{Bellhouse2017,Fritz2017,Gullieuszik2017,Poggianti2017b,Jaffe2018} 
we have used position-velocity phase-space diagrams of the cluster galaxies to infer the intensity of ram-pressure stripping, which, combined with the extent of the H$_{\alpha}$ emission in the disk, can be used to estimate the amount of gas removed \citep[see][for details on the method]{2015MNRAS.448.1715J, Jaffe2018}. One of the main caveats to this analysis is the uncertainty introduced by the projected distances and velocities that observations allow to measure. In this regard however, JO201 has a great advantage: given that it is being significantly stripped along the line of sight, we can assume that the measured projected velocity of the galaxy with respect to the cluster mean, is close to the true 3D velocity within the cluster. The observed projected clustercentric distance however is more likely to be an underestimation of the real 3D distance from the cluster centre. 

In GASP II we estimated the amount of stripping produced in JO201 based on its current location (projected distance 360~kpc from the cluster centre) and velocity (3363.7km~s$^{-1}$ or $3.4\times \sigma_{\rm cl}$) in the cluster (close to pericentre),
and compared it with the observed truncation radius ($r_t$) measured from the relative extent of the  H$\alpha$ emission relative to the stellar disk.

\begin{figure}\centering
\includegraphics[width=0.5\textwidth]{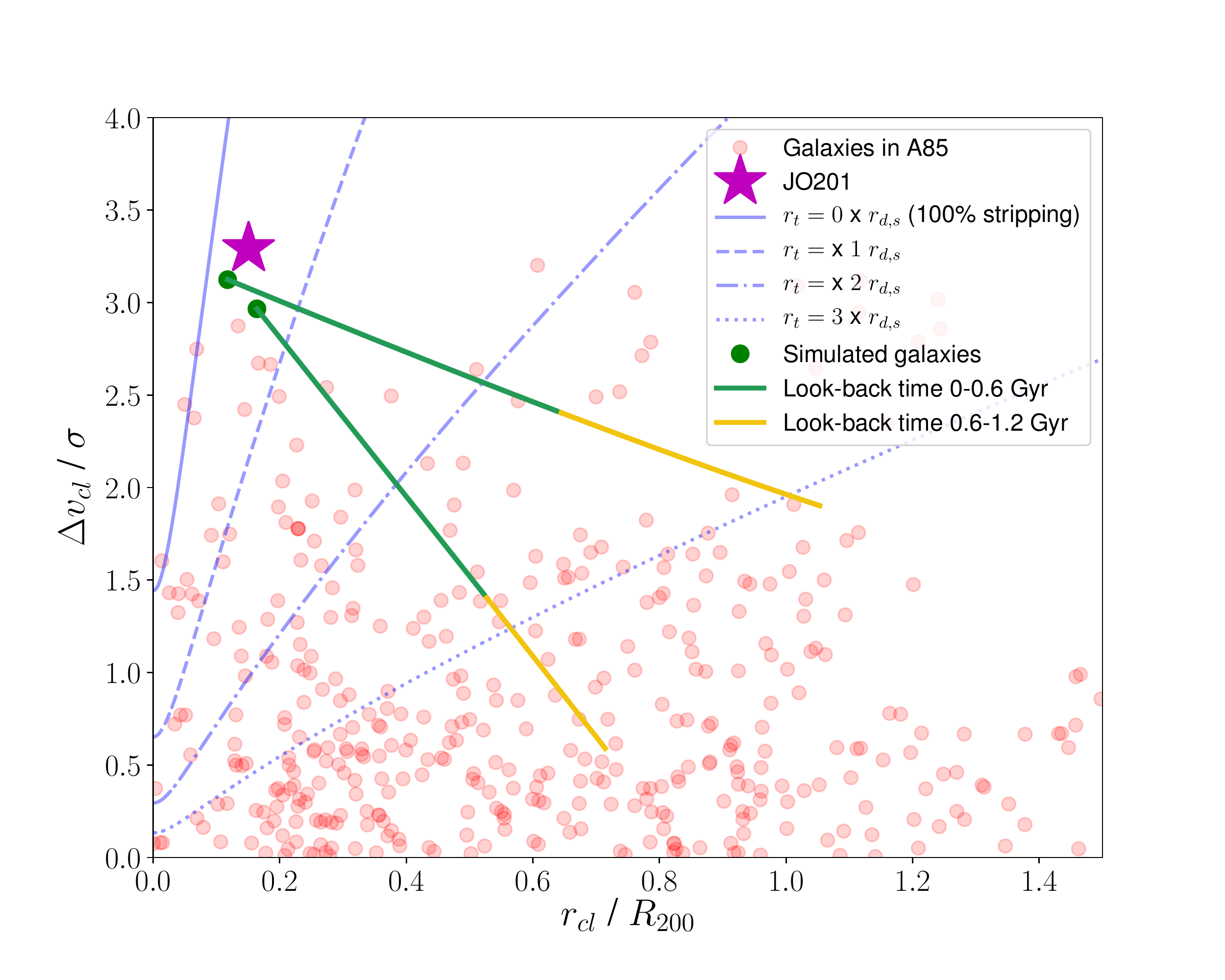}
\caption{Projected position-velocity phase-space diagram of all the galaxies in A85 (pink circles), including JO201 (star). 
Blue curves correspond to the estimated position and velocity of a JO201-like disk galaxy undergoing ram-pressure stripping by A85's ICM. 
The solid blue line that passes near the location of JO201 corresponds to total gas stripping via ram-pressure, using our revised derivation of $P_{\rm ram}$ and $\Pi_{\rm gal}$.
The dotted/dashed lines instead corresponds to an earlier stage of stripping, with a gas truncation radius $r_t =$ 1, 2, and 3~$\times R_{\rm d,stars}$. 
Finally, the coloured tracks correspond to orbits of simulated haloes in a massive cluster whose position in  phase-space at $z=0$ (green circles) is close to the location of JO201. The colours of the tracks indicate look-back time. 
} \label{figure:sims}
\end{figure}

We have re-done the ram-pressure calculation (Jaff\'e et al submitted) 
finding stronger ram-pressure stripping in JO201 than previously estimated.  
Our revised modelling\footnote{The corrected value of the ram-pressure at the position and velocity of JO201 is $P_{\rm ram}=2.7 \times 10^{-11}$~N~m$^{-2}$, and the anchoring pressure of the galaxy  $\Pi_{\rm gal}$ is $5.2 \times 10^{-11}$~N~m$^{-2}$ at the centre of the galaxy, using a revised disk scale-length of $r_{\rm d,stars}=2.5$~kpc,  which correspond to the mean  $R_{\rm d,s}$ of disk galaxies with a mass similar to JO201 \citep[see Fig. 1 in][]{Jaffe2018}. We chose to use this value rather than a fit to the light profile of JO201, because the latter could be affected by the jellyfish morphology.} 
suggests that the ram-pressure by the ICM should have almost completely removed gas from the galaxy at its current location in projected phase-space (see solid line in Figure~\ref{figure:sims}). This is somewhat consistent with our observations, which suggest there is still some gas left in the inner $\sim9$kpc of the disk (although disturbed in velocity with respect to the smooth rotation of the stellar disk; GASP II). 
The radius at which the gas becomes unbound from the stellar disk, and thus the total amount of gas lost, is difficult to measure (Gullieuzik et al. in prep), but from Fig. 15 in GASP II we estimate that the $H\alpha$ truncation radius in JO201 is $r_t \lesssim 9$~kpc, which translates to a total gas mass loss $>$ 40\%, assuming that $R_{\rm d,gas}=1.7 \times R_{\rm d,stars}$ prior to stripping.
The possible discrepancy between the observed $r_t$ and the predicted amount of stripping (Figure~\ref{figure:sims}) can be explained by the combination of an underestimated clustercentric distance (due to projection),  and the fact that \citet{1972ApJ...176....1G}'s prescription of ram-pressure assumes an instantaneous stripping of gas in an infinitely thin disk, a simplification that will over-predict the amount of gas stripping \citep[see Section 3.5 of][for a summary of all the caveats]{Jaffe2018}.

What we can be certain of from our observations is that the the galaxy is close to pericentre (and thus peak-stripping)  and that there has been a significant amount of gas stripping by ram-pressure which has affected even the inner-most parts of the disk.  

In order to estimate the timescale of stripping, it is necessary to know 
at which point in phase-space the galaxy started to lose gas.  
The answer is tightly linked to the orbital history of the galaxy, which we cannot measure directly but can infer from simulations.

We use N-body and hydrodynamical cosmological simulations from YZiCS \citep{Choi2017,Rhee2017} to track the orbits of haloes passing near the core of a massive ($9.23\times 10^{14} \mathrm{M}_\odot$) cluster at high speeds, like JO201 in A85. Using a random projection, we selected haloes whose end position in projected phase-space were the closest to the location of JO201, i.e. cluster-centric distances $r_{cl} < 0.5 \times R_{200}$, and high line-of-sight
velocity with respect to the mean velocity of the cluster: $\left| \Delta v_{cl} \right| > 2.75 \sigma$, where $\sigma$ is the velocity dispersion of the cluster. This selection yielded 2 simulated galaxies (see the circles in Figure~\ref{figure:sims}) which we tracked in  phase-space back in time until an epoch where the model galaxies had not yet experienced significant gas loss via stripping, i.e. to the right of the dotted blue line which corresponds to $\gtrsim$ 50\% of the total gas mass stripped. At this point the galaxy will have a gas deficiency of $\sim 0.3$, which is the threshold typically used to separate HI-rich galaxies from HI-deficient ones \citep{Haynes1984}. Although there is some uncertainty in the orbit of JO201 and the moment when stripping began, the model orbits indicate a timescale of $t_{\rm str} \sim 0.6-1.2$~Gyr from the beginning of significant gas stripping until the present epoch of ``peak" stripping near pericentre. This is shown by the yellow and green colours in the orbital tracks in Figure~\ref{figure:sims}. 

\subsection{Timescale of star-formation in the stripped tails}

An independent constraint of the timescale of stripping comes from the star-formation in the stripped tails, which presumably took place shortly after the stripping started 
\citep{Tonnesen2009,Tonnesen2010,Tonnesen2012}. 

An estimate of the average value of the stellar age was obtained by running {\sc sinopsis} on the integrated spectra of each knot. Two possible values of the stellar age can be defined: luminosity-weighted, and mass-weighted \citep{CidFernandes2005}, with the first definition being more directly observable and the second, while more physical, being more difficult to derive due to the strongly changing M/L as a function of the stellar population age.

The fact that the M/L of simple stellar populations is an increasing function of age, might pose a well known problem when using stellar population synthesis models to recover the properties of galaxies, in particular their star formation histories. In fact, the appearance of features due to a given stellar population in an observed spectrum depends both on the amount of stars and on their M/L. Since the stellar blobs we are analysing are formed in situ and outside the galaxy, the age of their stars must be younger than the age at which ram pressure started affecting the gas of the galaxy (Poggianti et al. 2018 in press). This is why when analysing the spectra of the blobs, we have set an upper limit of $2\times 10^9$ yr to the age of the stellar populations used for the spectral fitting. While being a rather conservative choice, this allows us to avoid spurious, artificial contributions in terms of stellar mass, and hence mass-weighted age, from older stars which, because of their very high M/L, would not provide a significant contribution to the observed spectrum.

Furthermore, once the star formation history was derived, we have calculated the contribution to the light of the observed spectra of each stellar population. If the contribution was less then 5\%, that population was not taken into account for the calculation of the mass-weighted age. We chose a threshold of 5\% because, for the typical S/N of the spectra we have analysed, this is the limit for which the fit is still acceptable.

\begin{figure}\centering
\includegraphics[width=0.5\textwidth]{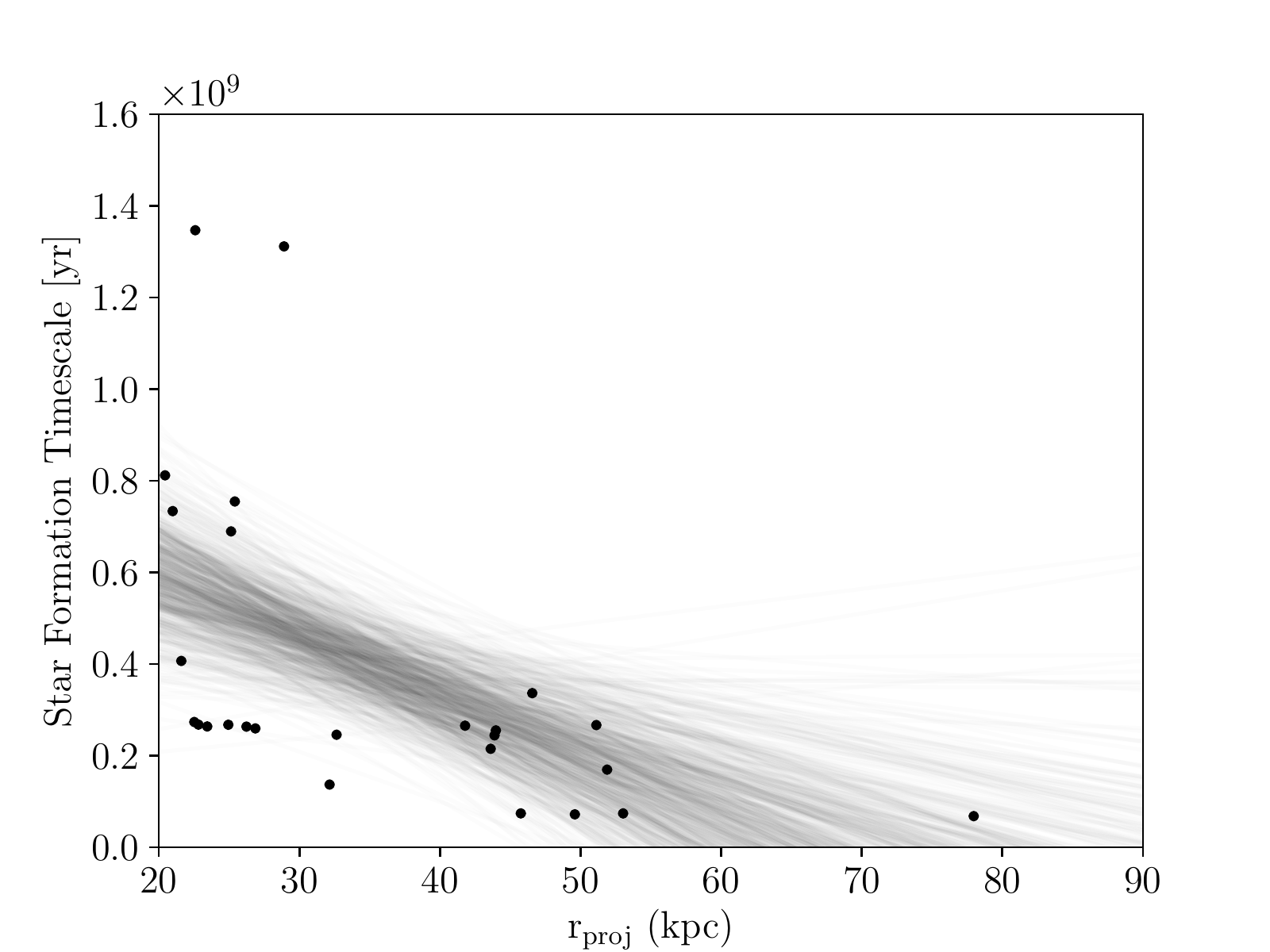}
\caption{Star Formation Timescale for knots $>$ 20kpc from JO201. Black points show the time since initial star formation began, approximated as $2 \times$ the mass weighted stellar age calculated using {\sc sinopsis}. A linear fit was applied using the python package linmix and transparent lines show samples from the posterior, indicating the scatter in the fit.}\label{figure:stellarageradius}
\end{figure}

The resulting age estimates are shown in Figure~\ref{figure:stellarageradius}, where the time since earliest star formation is shown for each knot against the projected radius for knots located in the tails outside the stellar disk ($r_{\rm proj}>20$kpc). Assuming a constant star formation history, the mass weighted stellar age will be the midpoint between the oldest and youngest stellar ages, therefore we approximate the time since earliest star formation as 2$\times$ the mass weighted stellar age calculated using {\sc sinopsis}.
%
Most of the star-forming knots in the tails have mass weighted stellar ages close to or less than 0.5 Gyr (between $10^{7}$ and $7 \times 10^8$ yr), suggesting that initial star formation began approximately 1 Gyr ago.




Interestingly, the figure also suggests (with a large scatter) a possible age gradient with the more distant knots having slightly younger stellar populations. 

This type of stellar age gradient along the tail has been observed in IC 3418 by \citet{Fumagalli2011,Kenney2014}, and also in RB 199 by \citet{Yoshida2008}, who attribute the age gradient to the movement of the stripped gas from which the knots form. The collapsed knots are left along the trajectory of the gas as it is stripped away from the galaxy. Once collapsed, the stars are no longer susceptible to ram-pressure, therefore knots that have collapsed earlier will both be preferentially situated closer to the galaxy and also contain older stars.

\subsection{Physical length of the stripped tails}

Given the extreme ram-pressure conditions that JO201 has been experiencing for the past $\sim 1$Gyr we expect the extent of the stripped tails to be large. 

For many jellyfish galaxies the physical size of the tails is not far from the measured projected size because the direction of motion is close to the plane of the sky. 
Because of the line-of-sight motion of JO201, the physical (unprojected) length of the stripped gas tails is not trivial to compute. However, we can obtain an estimate of the tail length by using a toy model to follow the motion and dynamics of a cloud of gas as it is accelerated away from the galaxy after being stripped by ram pressure. 
We model the infall of the galaxy into the potential well of a cluster with mass and density profiles derived from the X-Ray distribution of A85 \citep{Chen2007}. The galaxy begins at the virial radius of the cluster, 2.4Mpc, moving radially at 1500km~s$^{-1}$ (consistently with the orbits shown in figure~\ref{figure:sims}). It then accelerates as it falls into the potential well of the cluster. The cloud also falls into the cluster potential well, but its motion is additionally affected by acceleration due to ram pressure and, counteracting the ram pressure, acceleration from the potential well of J0201. The latter is modelled as a spherical NFW halo (M$_{200}$=1.0e12 M$_\odot$, c=15), and a 3D exponential disk (total mass=4.0e10 M$_\odot$, r$_{\rm exp}$=3.62 kpc, h$_{\rm z}$=0.25 kpc) using techniques described in \citet{Smith2015}, where all parameters are chosen to match the JO201 observations. We ran the model for a gas cloud with mass $4\times 10^5$ M$_\odot$ 
with a radius of 300~pc, which are typical values for HI clouds in the ISM (Vollmer 2001). The initial position of the cloud is in the outer disk of the galaxy at 4$\times R_{\rm d,stars}$, although our results are not strongly sensitive to this choice.
By the time the model galaxy reaches the current velocity and projected radius of JO201 within the cluster, the 
gas cloud has been stripped to 94kpc from the galaxy and has been accelerated until it has a velocity of 1200km/s relative to the galaxy, which is consistent with the largest velocity difference between the H$\alpha$ gas tails and the stars in JO201 (GASP II). 
We note however that the estimated length of the tails along the line of sight (and the velocity reached by the gas cloud) is sensitive to the mass of the cloud: A cloud of higher mass ($5.5\times 10^5$ M$_\odot$) reaches a shorter distance (68kpc) and relative velocity (1000km/s). 
In our model, the time required for the galaxy to reach this point since crossing the virial radius of the cluster is between 0.9-1.0Gyr, which lies within the range of estimates for the time since JO201 began stripping (Sec.~\ref{sec:timescale1}).

Our estimated upper limit to the line-of-sight length of the tails (94~kpc) is comparable to the longest tails observed in jellyfish moving on the plane of the sky such as JO206 \citep[90kpc,][]{Poggianti2017} and ESO 137-001 \citep[80kpc,][]{Sun2010}, which again suggests that the velocity vector of JO201 is dominated by the line of sight component.

Bringing together the results of the orbital analysis, star formation timescales and simulations, we find that JO201 began undergoing stripping between 0.6 and 1.2 Gyr ago as it approached the core of the massive cluster A85 at high speed along the line of sight. This resulted in tails up to 94kpc in length trailing behind the galaxy, which partially collapsed into knots within ~0.1 Gyr of initial stripping, forming stars within the past 0.1-0.7Gyr

\section{Conclusions}\label{sec:conclusions}

This work is part of a series of papers from the GASP (GAs Stripping Phenomena in galaxies with MUSE) \citep{Poggianti2017} survey of ``jellyfish'' galaxies, observed with the MUSE instrument at the VLT. In particular, this paper is a direct continuation of GASP II \citep{Bellhouse2017} which focused on the environment and  kinematic properties of JO201, a unique galaxy in the GASP sample subject to extreme ram-pressure conditions (i.e. very high speeds as it crosses the densest part of the ICM) in a trajectory close to the line of sight. In this paper, we expanded upon the previous study of the ionised gas and stellar kinematics with an analysis of the properties of the stripped gas, including the gas  metallicity, ionisation source and ionisation parameter. We also explored the stellar population ages, masses and star-formation across the galaxy, and build a toy model to estimate the timescale of the stripping event. 

We summarise our findings in the following:

\begin{itemize}

%
\item We detect several emission lines across the galaxy, including $\mathrm{H}\alpha, [\mathrm{NII}], [\mathrm{OIII}], \mathrm{H}\beta$ and $[\mathrm{SII}]$. The $\mathrm{H}\alpha$ emission notably extends well beyond the disk of the galaxy, while the distribution of other emission lines such as $[\mathrm{OIII}]$ are much more concentrated. The emission lines were used to determine the source of ionisation (via the BPT diagram), gas-phase metallicities and ionisation parameter across the galaxy disk and tails. 
\item The central region of the galaxy is dominated by broad AGN emission, surrounded by a ring of composite emission. The outer disk and tails are purely ionised by star-formation, concentrated in knots of low velocity dispersion gas.
\item The metallicity decreases in the stripped tails with distance from the galaxy disk. 
Such a metallicity gradient suggests that the outermost, lowest metallicity gas was stripped first, followed subsequently by gas of increasing metallicity as the stripping continued. This interpretation is supported by the increasing velocity offset of the gas (relative to the stars) at larger distances from the galaxy disk, shown in GASP II. 
%

\item The star-forming knots in the leading edge of the galaxy exhibit some of the highest SFRs in the galaxy, with values up to $6.0 \times 10^{-2}\mathrm{M}_\odot \mathrm{yr}^{-1} \mathrm{kpc}^{-2}$. These are likely to have formed as this part of the galaxy was compressed by the interaction with the ICM. In the tails the knots, which most likely formed in-situ from collapsed gas previously stripped from the disk, have lower SFR ($\sim3.1 \times 10^{-3}\mathrm{M}_\odot \mathrm{yr}^{-1} \mathrm{kpc}^{-2}$). 

\item Using {\sc sinopsis} to model the stellar population, we find that the oldest stars are only present in the disk, while the stellar component in the tails only began to form $\lesssim 0.7$Gyr ago. The most recent pockets of star-formation  
are located in the leading edge of the galaxy and in the knots within the tails.


\item We observe a gradient in the stellar age of the star forming knots, with the oldest knots residing closest to the galaxy. A possible explanation is that the knots collapse out of the stripped gas as it moves away from the galaxy, hence older knots collapsed earlier when the stripped gas was closer to the disk.

\item From orbital inference, we estimate that JO201 has been exposed to ram-pressure stripping for $\sim 1$ Gyr. 
This estimate is consistent with the 
time since initial star formation in the tails ($\lesssim 0.7$Gyr).

\item We use a model to estimate the physical (deprojected) length of the tails of gas stripped from JO201 along the line of sight. During infall of the model galaxy, we find that the stripping causes the tails to extend up to 94~kpc, which is comparable to tails observed in galaxies which are stripped predominantly on the plane of the sky.
\end{itemize}

Overall, in previous GASP studies \citep[including][]{Bellhouse2017,Poggianti2017b,George2018,Moretti2018} and the present paper we have exploited the rich MUSE data and other multi-wavelength observations on JO201, undoubtedly one the most spectacular and unique jellyfish galaxies in the GASP survey, to extract the physical properties in and around the galaxy and built a comprehensive view of the physics involved in the interaction between the galaxy and the ICM. 
In summary, all the different analyses done to this galaxy (in particular, the stellar and gas kinematics, ionisation sources, gas metallicity, and stellar population ages) individually and consistently point toward a scenario of incremental outside-in ram-pressure stripping of the gas during radial infall into the cluster, on a timescale $\sim1$ Gyr. The ram-pressure intensity is very high for this galaxy compared with other jellyfish galaxies because it is travelling close to face-on at supersonic speeds though the hot and dense medium of a massive cluster, and is in a moment in its (radial) orbit close to pericentric passage. The ram-pressure wind is not only able to remove a significant amount of gas  from the disk 
but it is also likely to be pushing some of the gas towards the centre, fuelling the supermassive black hole. At the same time, the gas in the leading edge of the galaxy and in the stripped tails is compressed, giving rise to new episodes of star-formation, which is concentrated into knots. 

Moreover, our estimated stripping timescale and ages of the stellar populations in the star-forming knots outside the galaxy 
confirm that some of the knots in the tails must have formed in-situ, consistent with hydrodynamical models of RPS and observations of this galaxy at other wavelengths \citep{Tonnesen2009,Tonnesen2010,Tonnesen2012,George2018,Moretti2018} 

%
Future studies will make use of the large sample of jellyfish galaxies observed by the GASP survey to quantify, in a statistical way, different aspects of the consequences of RPS, such as the contribution of stripped mass from galaxies into the ICM, and the impact of RPS on the internal properties of galaxies and their subsequent evolution.



\section*{Acknowledgements}

Based on observations collected at the European Organisation for Astronomical Research in the Southern Hemisphere under ESO programme 196.B-0578. 
Y.~J. acknowledges support from CONICYT PAI (Concurso Nacional de Inserci\'on en la Academia 2017) No. 79170132. We acknowledge financial support from PRIN-SKA 2017 (PI Hunt). 
We would like to thank Hoseung Choi and Sukyoung Yi for providing access to YZiCS simulation data.

\bibliographystyle{mnras}

\clearpage
\appendix

\section{Star Forming Knot Properties}

\begin{figure*}\centering
\includegraphics[width=0.32\textwidth]{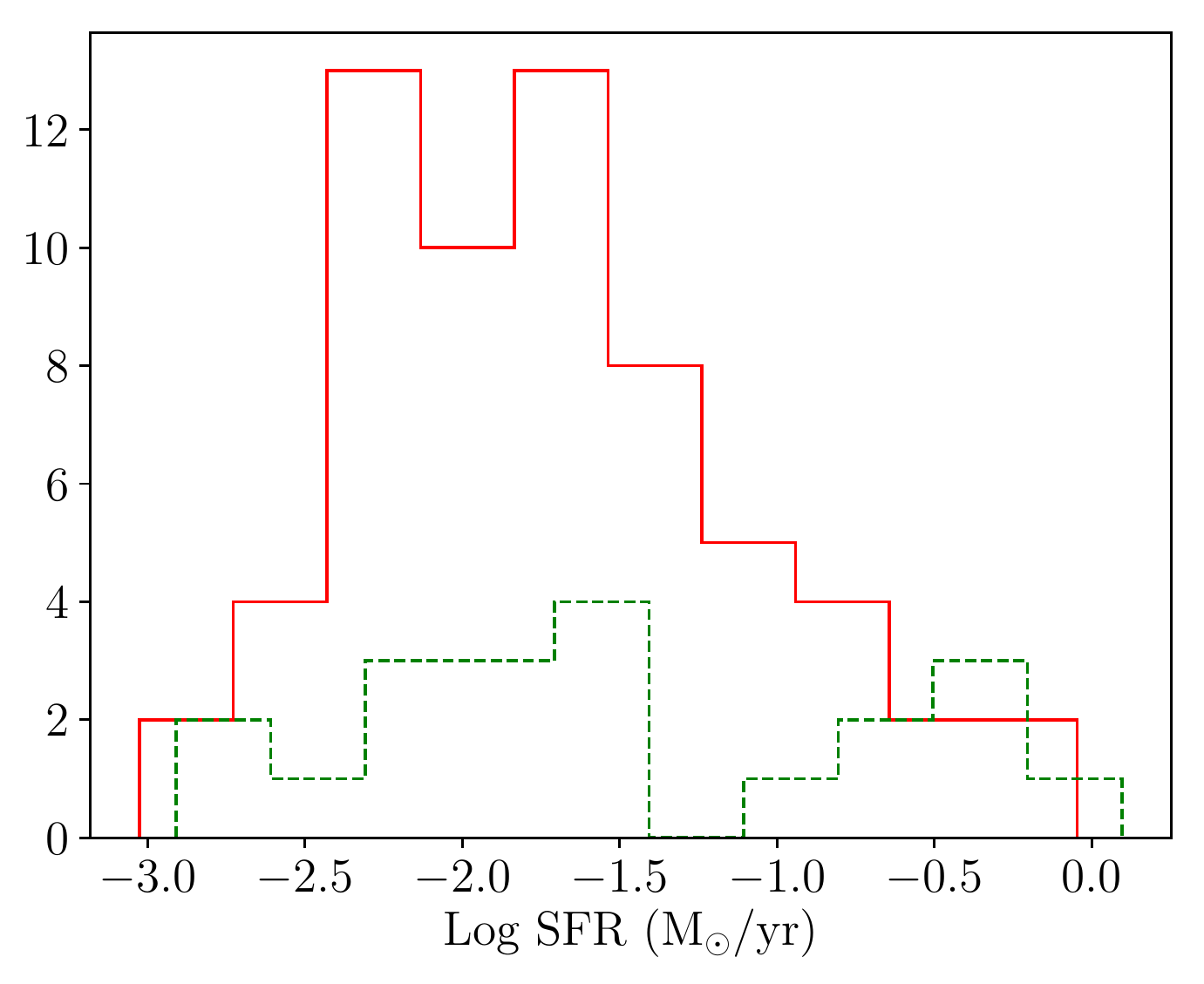}
\includegraphics[width=0.32\textwidth]{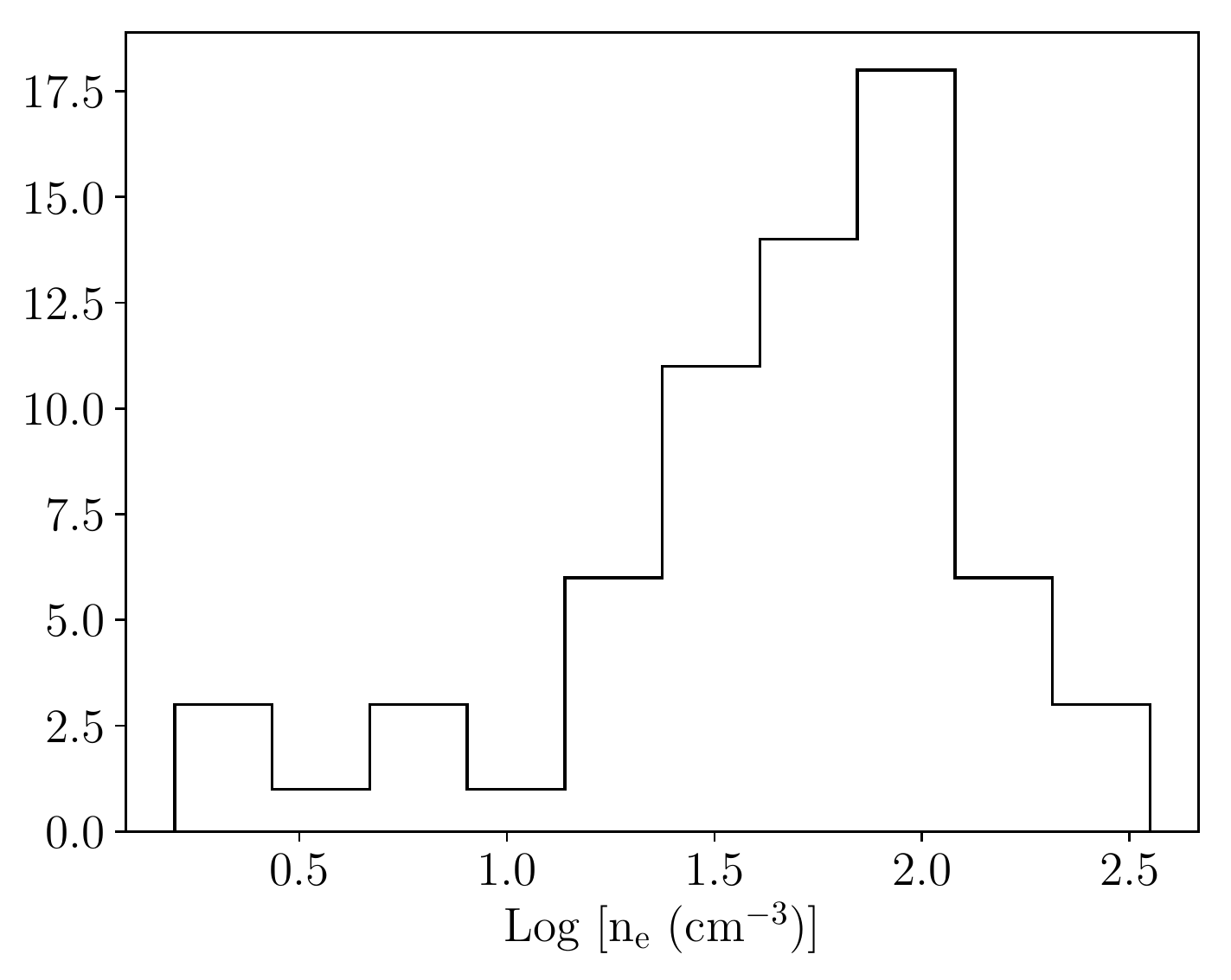}
\includegraphics[width=0.32\textwidth]{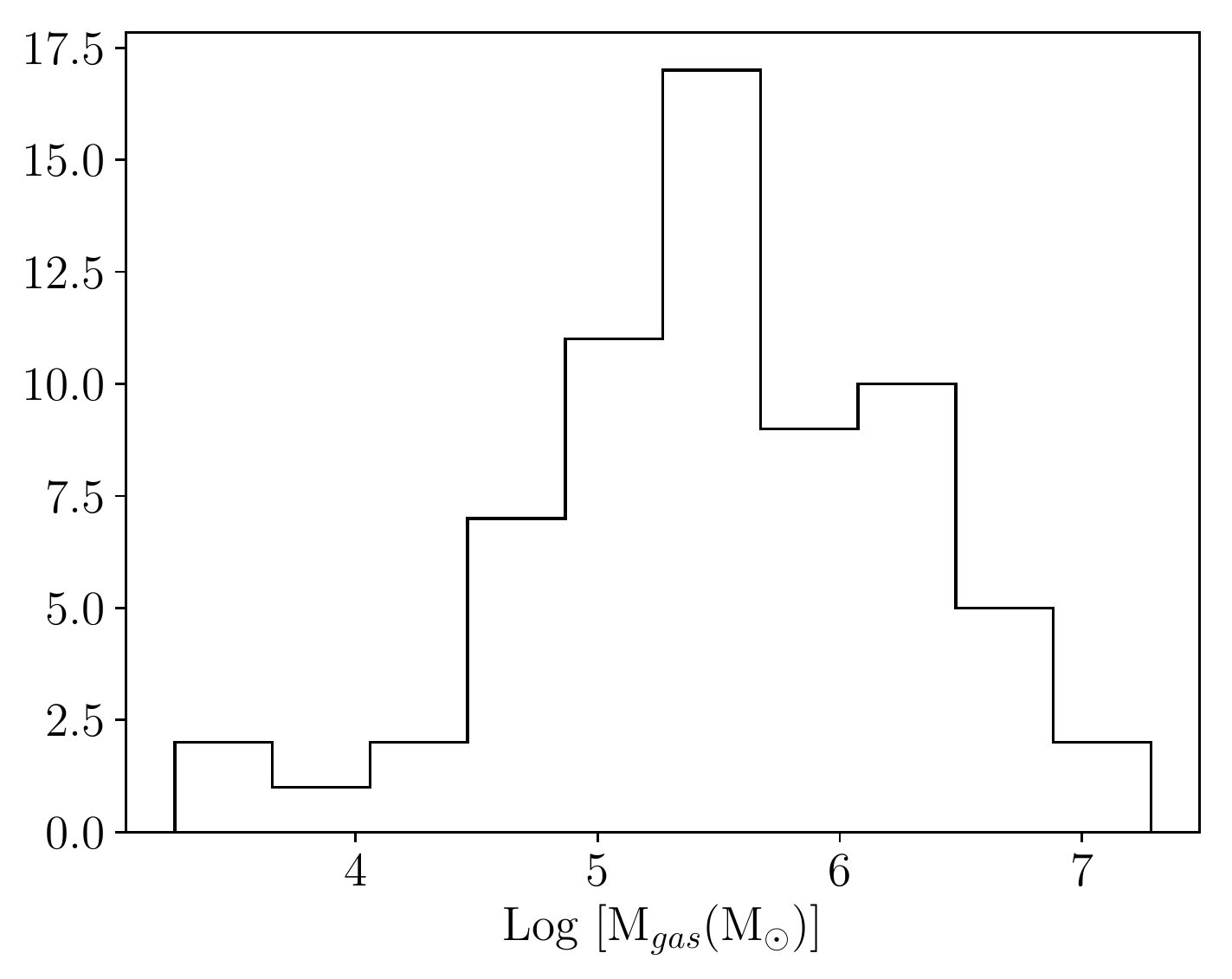}
\caption{\textit{Left:} Distribution of SFR for star-forming (red, solid) and composite (green, dashed) knots. The central knot, dominated by AGN emission, has been excluded. \textit{Middle:} Gas density distribution of individual knots \textit{Right:} Distribution of ionised gas masses within knots.}\label{figure:hist_blobs}
\end{figure*}

Figure~\ref{figure:hist_blobs} shows the star-formation, gas density and gas mass calculated as in GASP I, for the knots in JO201 discussed in section \ref{sec:knots}.
The first plot shows the distribution of star-formation rate (SFR) for knots below (red) and above (green) the \citet{Kauffmann2003} line of the BPT diagram. The star-formation rates are calculated using the dust extinction and stellar absorption corrected H$\alpha$ luminosity in the \citep{1998ARA&A..36..189K} method transformed to a Chabrier IMF, as described in GASP I. We note that the SFRs obtained here are slightly lower (by 0.1-0.5M$_\odot \mathrm{yr}^{-1}$) than those computed from far ultraviolet emission by \citet{George2018} (using a Salpeter IMF), which could be attributed to the uncertainty in the dust extinction correction.

The second plot in Figure~\ref{figure:hist_blobs} shows the density of the ionised gas calculated using the ratio of the [SII] 6716 and [SII] 6732 lines. The calculation is made using the \citet{Proxauf2014} calibration for $T=10$,$000K$.

The third plot of Figure~\ref{figure:hist_blobs} shows the distribution of the ionised gas masses of the knots, computed from the H$\alpha$ luminosities following equation 13.7 in \citet{OsterbrockFerland} as in \citet{Boselli2016} and \citet{Poggianti2017}. Case B recombination is assumed with $n=10$,$000cm^{-3}$ and $T=10$,$000K$, and it is assumed that the gas is fully ionised.

In comparison, JO201 has similar peaks and distributions in each plot as JO206, shown in \citet{Poggianti2017}, and also JO204 \citep{Gullieuszik2017}, with slightly higher SFRs in general in the knots and slightly higher gas densities.

\end{document}